\documentclass[12pt]{iopart}

\usepackage{graphicx}

\begin{document}

\title[Neutron electromagnetic structure]{The chiral structure of the neutron as revealed in electron and photon scattering}

\author{D.\ R.\ Phillips}

\address{Department of Physics and Astronomy, Ohio University, Athens, OH 45701, USA}
\ead{phillips@phy.ohiou.edu}
\begin{abstract}
It is a consequence of QCD's spontaneously broken chiral symmetry that the response of the neutron to 
long wavelength probes will be driven by excitation of its pion cloud. In this article I discuss 
manifestations of this in electromagnetic form factors and Compton scattering. Particular attention is given to the interplay between this physics and shorter-distance components of the neutron's electromagnetic structure. 
\end{abstract}

\maketitle

\section{Introduction}

In the limit of massless up and down quarks the QCD Lagrangian has an $SU(2)_L \times SU(2)_R$ symmetry. We know from lattice simulations that the vacuum of QCD in this limit does not share this symmetry of the Lagrangian, 
with the axial combination of $SU(2)$'s being spontaneously broken~\cite{ChiSBoutnoChiSBin}. This confirms the key elements of the picture postulated by Nambu in talks and papers of 1960--1, work for which he was awarded the 2008 Nobel Prize in Physics~\cite{Na60,NJL61}. An immediate consequence of the spontaneous breaking of chiral symmetry is that there will be three Goldstone bosons: the $\pi^+$, $\pi^0$, and $\pi^-$. Since the up and down quarks are not exactly massless these pions are also not massless, but, to the extent that $m_u$ and $m_d$ are small compared to $\Lambda_{QCD}$, the pions will be much lighter than all other hadronic degrees of freedom.

This picture is systematized in an effective field theory (EFT) called `chiral perturbation theory' ($\chi$PT)~\cite{Weinberg79,GL82,Gasser:1983yg,Gasser:1984gg}. Chiral perturbation theory encodes the  $SU(2)_L \times SU(2)_R$ symmetry of QCD, and the pattern of its breaking, in a Lagrangian containing nucleon and pion fields. It includes all possible interactions consistent with the symmetry. These are organized in a derivative expansion, together with an expansion in powers of $m_q$ for the symmetry-breaking operators. The EFT can be used to calculate both tree and loop diagrams, which leads to a perturbation expansion for scattering amplitudes in powers of 
\begin{equation}
P \equiv \frac{p,m_\pi}{\Lambda_\chi}.
\label{eq:P}
\end{equation}
Here $\Lambda_\chi \sim 4 \pi f_\pi \approx 1200$ MeV (with $f_\pi \approx 93$ MeV the pion-decay constant) is the scale that emerges from loop calculations~\cite{MG83} as controlling the convergence of the expansion for meson-meson interactions. If we consider threshold processes, and so $p=0$, this is an expansion in powers of $m_\pi$, and so can be considered an expansion around the chiral limit of QCD, where $m_\pi=0$. In the meson sector the theory is mature, and calculations are routinely performed to two loops. For $\pi \pi$ scattering the results are extremely accurate, reaching the sub-1\% level. (For a recent review see \cite{Bi07}.) 

Adding baryons to the theory as static sources facilitates a power counting in which each diagram has a definite order in the $P$ expansion~\cite{JM91}. The effects of baryon recoil constitute corrections to the static-source picture, and they can be systematically included in the Lagrangian via an expansion in powers of $p/M$~\cite{JM91,Be92}. The result is known as ``heavy-baryon chiral perturbation theory" (HB$\chi$PT). HB$\chi$PT for nucleons and pions is summarized in Section~\ref{sec-chipt1N}. There has been a recent flurry of activity devoted to baryon $\chi$PT calculations that do not invoke this $p/M$ expansion~\cite{BL00,Fu03}, but here I will be mainly using the heavy-baryon variant of the theory to describe neutron structure. Electromagnetic interactions are easily included in this theory, via minimal substitution and the addition of terms that are gauge invariant by themselves (e.g. an operator corresponding to magnetic fields coupling to the neutron anomalous magnetic moment). In the usual $\chi$PT counting the electron charge $e$ then counts as one power of $P$. For comprehensive reviews of $\chi$PT in the baryon sector (both heavy-baryon and ``relativistic" formulations) see Refs.~\cite{BKM,Sc02,BM07,Be08}. 

The standard $\chi$PT Lagrangian includes only nucleons and pions, and so its breakdown scale is not $4 \pi f_\pi$. Instead it is set by the lightest hadronic degree of freedom not explicitly included in the theory, which in most reactions is the Delta(1232). So, unless the $\chi$PT Lagrangian is extended to include Delta degrees of freedom,  the scale in the denominator in Eq.~(\ref{eq:P}) is not $\Lambda_\chi$, but is instead $M_\Delta - M_N \approx 290$ MeV. While there has been much progress in this direction~\cite{JM91,He97,He98,PP03,Hi04,LP09}, for most of this article I will regard the Delta-nucleon mass difference as a high-energy scale. 

My focus here is thus on the {\it chiral} structure of the neutron, as revealed by electromagnetic probes. I therefore report on computations of electron and photon scattering, with the main emphasis on results for these reactions obtained in HB$\chi$PT.  I will be particularly concerned with the interplay between pion-cloud mechanisms and the role of higher-energy excitations. The ability to vary photon energy, $\omega$, in the case of Compton scattering, and  three-momentum transfer, $\bf q$, in the case of electron scattering, allows us to map out this interplay over a significant kinematic domain. It also provides interesting data on the breakdown of $\chi$PT, since comparison of its predictions with experimental data over a range of $\omega$ and $|{\bf q}|$ can reveal where the theory is failing, and what additional degrees of freedom are needed in order to extend its radius of convergence. 

But, practically speaking, these aspects of neutron structure cannot be examined separately from the chiral structure of nuclei. Neutrons cannot be collected in sufficient numbers to make electromagnetic scattering reactions on free neutrons an experimental reality. Hence everything we know about neutron electromagnetic structure is inferred from more complex processes, e.g. experiments on light nuclei. In order to extract information on neutron structure from such experiments an understanding of the multi-nucleon effects that are present in the data is also necessary.
 
Developments over the past 15 years have also allowed $\chi$PT to provide a systematic approach to multi-nucleon problems. Weinberg's seminal papers on $\chi$PT expansions for the nuclear potential~\cite{We90,We91} and the operators governing the interaction of external probes with the nucleus~\cite{We92} facilitate a parallel and consistent $\chi$PT treatment of neutron structure and electromagnetic reactions in two- (and three- and \ldots) nucleon systems. In Section~\ref{sec-chipt2N} I will give a brief explanation of how $\chi$PT enables the consistent treatment of the operators governing processes in single- and multi-nucleon systems. Electron-neutron scattering and $\gamma n \rightarrow \gamma n$ will then be discussed in Sections~\ref{sec-ffs} and \ref{sec-compton} respectively. In this article I only have space to discuss those two processes in detail, so in the final section, Section~\ref{sec-conc}, I briefly mention some other reactions that probe the same physics: virtual Compton scattering, as well as pion photo- and electro-production near its threshold. I also offer some concluding thoughts on the neutron's chiral structure. 

\section{Chiral perturbation theory for nucleons and pions}

\label{sec-chipt1N}

In this section I provide a brief summary of the aspects of chiral perturbation theory ($\chi$PT) that are relevant for the subsequent presentation.
The discussion here follows the notation and conventions of Ref.~\cite{BKM}. In particular, I will use the `heavy-baryon' formulation of $\chi$PT, in which a
$p/M$ expansion is made alongside the expansion in powers of $p/\Lambda_\chi$. This expansion is not essential to the efficacy of $\chi$PT, but since $M \sim \Lambda_\chi$
it does not have a deleterious effect on its convergence---apart from a few specific cases that involve proximity to kinematic thresholds.

$\chi$PT is an effective field theory of QCD, that encodes the symmetries of the underlying theory which are pertinent to the regime where energies are of the order of Goldstone boson (here, pion) masses. As such the chiral Lagrangian is formulated in terms of objects with straightforward transformation properties under the chiral group $SU(2)_L \times SU(2)_R$. For details on the transformation properties of the building blocks of the chiral Lagrangian the reader is referred to Refs.~\cite{BKM,SS05}. Here we choose a particular representation for the pion field, and summarize the building blocks as:
\begin{equation}
u=\exp\left(i \frac{\tau \cdot {\bf \pi}}{2 f_\pi}\right); \quad U=u^2;
\end{equation}
where ${\bf \pi}$ is the pion iso-triplet. From these we construct the chiral covariant derivative of the nucleon field, $D_\mu$, and the axial-vector object with one derivative, $u_\mu$:
\begin{eqnarray}
D_\mu \psi&=&\partial_\mu \psi + \Gamma_\mu \psi; \\
\Gamma_\mu&=&\frac{1}{2}\left[u^\dagger \partial_\mu u +u  \partial_\mu u^\dagger  - i e A_\mu \left(u^\dagger {\cal Z} u + u {\cal Z} u^\dagger\right)\right];\\
u_\mu&=&i \left[u^\dagger \partial_\mu u - u \partial_\mu u^\dagger - i e A_\mu \left(u^\dagger {\cal Z} u - u {\cal Z} u^\dagger\right)\right],
\end{eqnarray}
with ${\cal Z}={1 \over 2}(1 + \tau_3)$. Here we have specialized to  the external-field case of interest for this paper: $A_\mu$ is an (external) photon field and there is no external axial field. Note that the vector field $\Gamma_\mu$ contains only even powers of the pion field, and $u_\mu$ contains only odd powers of it. The nucleon field $\psi$ is the usual iso-double. Both components of $\psi$ are Dirac fields. 

The leading-order Lagrangian contains all possible terms with zero and one derivative and is:
\begin{equation}
 {\cal L}_{\pi N}^{(1)}=\bar{\psi} \left(i \not \! \! D - M + \frac{g_A}{2} \gamma^\mu u_\mu \gamma_5\right)\psi.
 \label{eq:LpiNrel1}
 \end{equation}
 Strictly speaking $M$ and $g_A$ here are the mass and axial coupling of the nucleon in the chiral limit, although the distinction between their values at zero quark mass and the physical ones will not be important for any of the discussions here. This Lagrangian, supplemented by the standard leading-order meson-meson one:
 \begin{equation}
 {\cal L}_{\pi \pi}^{(2)}=\frac{f_\pi^2}{4} [{\rm Tr}(D_\mu U^\dagger D^\mu U) - 2 B  {\rm Tr}({\cal M}(U + U^\dagger))]
 \label{eq:Lpipi2}
 \end{equation}
 with $D_\mu U=\partial_\mu U - i e A_\mu [\frac{1}{2} \tau_3,U]$, is all one needs to calculate the leading-order loop effects in chiral perturbation theory.
 
Such calculations are rendered more straightforward, and more transparent, if one removes the large energy scale associated with the nucleon mass. This energy is not available for particle production, because of baryon-number conservation. Hence it plays no role in the single-nucleon sector, and can be eliminated from consideration by what amounts to a different choice of the zero of energy. This is done explicitly by use of a field redefinition. As first pointed out by Jenkins and Manohar~\cite{JM91}, the field $\psi$ can be re-expressed in terms of a ``heavy-baryon" field $\psi_{HB}$, that has the straight-line propagation of a nucleon of mass $M$ removed from its dynamics:
 \begin{equation}
 \psi(x) \equiv \psi_{NR}(x) \exp(-i M v \cdot x).
 \label{eq:HB}
 \end{equation}
If the nucleon is at rest then the choice $v=(1,\bf{0})$ is appropriate, although physical results should be independent of the choice of $v$. We also define the analogue of the non-relativistic spin
\begin{equation}
S_\mu=\frac{i}{2} \gamma_5 \sigma_{\mu \nu} v^\nu.
\end{equation}
If we choose $v=(1,\bf{0})$ and consider the case of four space-time dimensions then the spatial components of $S$ obey the algebra:
\begin{equation}
\{S^i,S^j\}=\frac{1}{2} \delta_{ij}; \quad [S^i,S^j]=i \epsilon_{ijk} S^k.
\end{equation}
Thus a representation for them is provided by the assignment $S^i=\Sigma^i/2$, with $\Sigma^i$ the block-diagonal Dirac matrix constructed out of the Pauli matrices $\sigma^i$. 

Here we focus on the leading pieces of the heavy-baryon Lagrangian, i.e. the dominant terms that emerge from Eq.~(\ref{eq:LpiNrel1}) in a $p/M$ expansion (with $p$ the three-momentum of the nucleon state). The key observation is that the operator structures in Eq.~(\ref{eq:LpiNrel1}) do not couple the
upper and lower components of $\psi_B$ to one another. Hence, 
the leading behavior of all Dirac bilinears in the $p/M$ expansion can be obtained by assuming that $\psi_B=H$ with $H$ obeying $\mbox{$\not \! v$} H=H$, i.e. assuming $\psi_B$ contains only two degrees of freedom. We identify these as the degrees of freedom associated with the usual non-relativistic spin operator.

When written in terms of bilinears of $H$, the operator structures in Eq.~(\ref{eq:LpiNrel1}) involve 
only the four-vectors $v$ and $S$, since:
\begin{eqnarray}
\bar{H} \gamma_\mu H&=&[v_\mu] \bar{H} H \label{eq:id1}\\
\bar{H} \gamma_\mu  \gamma_5 H&=&2 \bar{H} S_\mu H. \label{eq:id2}
\end{eqnarray}
These identities, when applied to the version of Eq.~(\ref{eq:LpiNrel1}) with Eq.~(\ref{eq:HB}) inserted into it, produce the standard leading-order heavy-baryon Lagrangian:
\begin{equation}
{\cal L}_{\pi N}^{(1)}=\bar{H} \left[i(v \cdot D) + g_A u_\mu S^\mu\right] H.
\label{eq:LpiN1}
\end{equation}

The structure of the propagator for the $H$ field can be understood by writing 
the $\psi$ field's total four-momentum, $k$, as $k=Mv + p$. Then, in terms of the small residual (``off-shell") momentum $p$, the Dirac propagator may be written:
\begin{equation}
S=\frac{i}{\not \! k - M} =i\frac{M \not \! v + \not \! p + M}{2 M v \cdot p + p^2} =\frac{i}{v \cdot p} \frac{1 + \not \! v}{2} + \ldots.
\label{eq:HBprop}
\end{equation}
The first term is this expansion is then the propagator for the $H$ field. 
In a typical $\chi$PT loop graph for a single-baryon process the baryon emits and absorbs a pion, and so it is easy to see that the loop graph will be dominated by contributions from momenta for which all four components of $p$ are of order $m_\pi$. The $\ldots$ terms in Eq.~(\ref{eq:HBprop}) are then suppressed by $m_\pi/M$. Exceptions to this rule occur in the vicinity of kinematic thresholds, but, as long as such thresholds are not nearby, the leading effects of the loop graphs in $\chi$PT can be obtained by replacing the full baryon propagator by the heavy-baryon one (\ref{eq:HBprop}).

The corrections of higher order in $1/M$ can be systematically derived by writing 
\begin{equation}
\psi_B=H + h
\end{equation}
with 
\begin{equation}
\not \!  v H=H; \qquad \not \! v h=-h,
\end{equation}
and then calculating the Lagrangian for the field $H$ that results from eliminating the degrees of freedom associated with $h$. This treatment of all $1/M$ corrections to the leading-order heavy-baryon Lagrangian (\ref{eq:LpiN1}) has been carried out by Bernard, Kaiser, Kambor, and Mei\ss ner to $O(P^2)$~\cite{Be92}, Fettes, Mei\ss ner and Steininger to $O(P^3)$~\cite{FMS}, and Fettes, Mei\ss ner, Moj\v{z}i\v{s}, and Steininger to $O(P^4)$~\cite{FMMS}. At fourth order a large number of operators in the non-relativistic theory of which (\ref{eq:LpiN1}) is the leading Lagrangian have coefficients that are fixed by the requirement that S-matrix elements at low energies agree with the S-matrix elements of the relativistic theory.

Such considerations regarding the low-energy consequences of Lorentz invariance generate the first three terms in the second-order Lagrangian (for further details the reader is referred to Ref.~\cite{BKM}). But, since this is an effective theory, the second-order Lagrangian contains all operators that are consistent with the symmetries and have two derivatives or one power of the quark-mass matrix (since $m_q \sim m_\pi^2$ and $p \sim m_\pi$). There are six such independent operators in the isospin-symmetric limit $m_u=m_d$. The complete second-order Lagrangian is then~\cite{BKM}:
\begin{eqnarray}
{\cal L}_{\pi N}^{(2)}&=&\bar{H} \left\{\frac{1}{2 M} [(v \cdot D)^2 - D \cdot D)] - \frac{i g_A}{2 M} \{S \cdot D, v \cdot u\} + c_1 {\rm Tr} \chi_+\right.\nonumber\\
&&+ \left(c_2 - \frac{g_A^2}{8 M}\right) (v \cdot u)^2 + c_3 u \cdot u + \left(c_4 + \frac{1}{4 M}\right)[S_\mu,S_\nu] u^\mu u^\nu \nonumber\\
&& - \left.\frac{i}{4M} [S_\mu, S_\nu] \left[(1 + \kappa_v) f_{\mu \nu}^+ + \frac{1}{2} (\kappa_s - \kappa_v) {\rm Tr}(f_{\mu \nu}^+)\right] \right\} H,
\label{eq:LpiN2}
\end{eqnarray}
with 
\begin{equation}
f_{\mu \nu}^+=e F_{\mu \nu} \left(u^\dagger {\cal Z} u + u {\cal Z} u^\dagger\right),
\end{equation}
the electromagnetic-field-strength tensor (containing both isovector and isoscalar parts) that transforms in the appropriate way under the chiral group.
In this formulation of hadron dynamics the values of two of the low-energy constants  (LECs) that appear in (\ref{eq:LpiN2}) have already been set to the measured isoscalar and isovector nucleon magnetic moments, $\kappa_s$ and $\kappa_v$~\footnote{Strictly speaking, it is the chiral-limit magnetic moments that enter here, but the difference is a fourth-order effect.}. The other LECs, $c_1$--$c_4$, encode the effects of more massive baryon and meson resonances ($\Delta(1232)$, Roper, $\rho$-meson, \ldots) on low-energy dynamics. 

The terms listed in Eqs.~(\ref{eq:Lpipi2}), (\ref{eq:LpiN1}), and (\ref{eq:LpiN2}) are all that is required for computation of complete one-loop effects in HB$\chi$PT. These loops are divergent, but the degree of divergence is mandated by the power of momentum carried by the vertices, and the behavior of the heavy-baryon and meson propagators. Dimensional analysis, combined with the usual topological identities for Feynman graphs, shows that 
a $\chi$PT graph with $L$ loops, $N_d^{\pi \pi}$ vertices from the $\pi \pi$ Lagrangian of dimension $d$, and $N_d^{\pi N}$ vertices from the $\pi$N Lagrangian of dimension $d$ has a total degree of divergence:
\begin{equation}
D=2L + 2 - A + \sum_d (d-2) N_d^{\pi \pi}+ \sum_d (d-1) N_d^{\pi N},
\label{eq:chiptpc}
\end{equation}
with $A$ the number of nucleons. 

This calculation of the degree of divergence of the graph is important, because, after renormalization is performed, the finite part of the graph must scale as $(p,m_\pi)^D$. Therefore, provided $p \sim m_\pi$, the computation of $D$ tells us the order in $\chi$PT at which the graph contributes. If renormalization is performed using dimensional regularization and a mass-independent subtraction scheme then the effects of any particular HB$\chi$PT graph occur at one and only one chiral order $D$. This is an advantage of the heavy-baryon formulation: each graph scales precisely as $P^D$, and so has a definite order in the chiral counting. Specifically, Eq.~(\ref{eq:chiptpc}) says that one-loop graphs with vertices from ${\cal L}_{\pi N}^{(1)}$ give pieces of the quantum-mechanical amplitude with $D=3$, while graphs with one insertion from
${\cal L}_{\pi N}^{(2)}$ give effects with $D=4$. Furthermore, since ${\cal L}_{\pi \pi}$ starts at $d=2$, and ${\cal L}_{\pi N}$ at $d=1$, the lowest order at which a two-loop graph can contribute is $D=5$. Thus one-loop graphs with vertices from ${\cal L}_{\pi N}^{(1)}$ and either zero or one insertion(s) from ${\cal L}_{\pi N}^{(2)}$---together with appropriate tree-level contributions---give the complete $\chi$PT result up to $O(P^4)$.

Of course, the divergent loops at $O(P^3)$ and $O(P^4)$ need to be renormalized, and this is done with counterterms that occur in ${\cal L}_{\pi N}^{(3)}$ and ${\cal L}_{\pi N}^{(4)}$. Much effort has gone into finding a complete, minimal Lagrangian at these orders (see, e.g. Refs.~\cite{FMMS,EM96}). I will not attempt to do justice to that effort here, since  only a few of the many contact interactions that occur in ${\cal L}_{\pi N}^{(3)}$ and ${\cal L}_{\pi N}^{(4)}$ enter the results presented below. I want to stress, though, that it is through such counterterms that hadronic states with excitation energies $> m_\pi$ affect $\chi$PT's predictions. These states therefore produce effects that are slowly varying with respect to the external kinematic parameter $|{\bf q}|$ or $\omega$, while the rapid variation of measurable quantities with $|{\bf q}|$ and $\omega$ is provided by pion-cloud physics. This interplay of the chiral physics with the higher-energy hadron dynamics is captured in $\chi$PT, without the need for any assumptions about the details of that dynamics.

\section{Chiral perturbation theory in multi-nucleon systems}

\label{sec-chipt2N}

The previous section discussed chiral perturbation theory for the single-nucleon system. In subsequent sections we will examine the predictions of this theory for photon and electron scattering on the neutron. However,``experimental" data on these reactions can only be obtained using nuclear targets. In analyzing these data it is beneficial to treat the interactions that bind the nucleus within the same framework as that used to describe the chiral structure of the nucleon itself. This can be done in chiral perturbation theory, which not only provides a means to compute the low-energy interaction of photons and pions with nucleons, but also is used to calculate the forces between nucleons. This is the framework that will be used, e.g. in Section~\ref{sec-compton}, to compute photon interactions with multi-nucleon systems, and we provide a brief summary of it here.

The first attempt in this direction was due to Weinberg~\cite{We90,We91}. Weinberg pointed out that HB$\chi$PT could not be employed directly to compute the nucleon-nucleon amplitude, because a nucleon in a typical nuclear bound state has 
\begin{equation}
p_0 \sim {\bf p}^2/M; \quad |{\bf p}| \sim m_\pi.
\label{eq:nucscaling}
\end{equation}
Thus not all components of the (residual) nucleon four-momentum $p$ are order $m_\pi$, in contrast to the case inside a typical HB$\chi$PT loop graph. Because of the scaling (\ref{eq:nucscaling}) a different organization of the Lagrangian (\ref{eq:LpiN1}) plus (\ref{eq:LpiN2}) is required when treating NN interactions. The single-nucleon propagator becomes:
\begin{equation}
S(p)=\frac{i}{v \cdot p - \frac{(v \cdot p)^2 - p^2}{2 M}} \stackrel{v=(1,{\bf 0})}{\longrightarrow} \frac{i}{p^0 - \frac{{\bf p}^2}{2M}}, 
\end{equation}
i.e. the usual non-relativistic propagator for a free particle of mass $M$. 
Weinberg further observed that in the regime (\ref{eq:nucscaling}) this non-relativistic propagator scales as $M/m_\pi^2$, which invalidates the degree-of-divergence formula (\ref{eq:chiptpc}). 

However, if we consider only diagrams for $NN \rightarrow NN$ that do not contain an intermediate two-nucleon state---the so-called `two-particle irreducible' (2PI) graphs---then all single-nucleon propagators have $p_0 \sim m_\pi$, and
Eq.~(\ref{eq:chiptpc}) can be used. The goal thus becoomes to compute the 2PI NN amplitude up to a fixed chiral dimension $D$.
 In order to do this we must supplement ${\cal L}_{\pi N}$ and ${\cal L}_{\pi \pi}$ by a nucleon-nucleon Lagrangian that contains the effects of short-distance ($r \ll 1/m_\pi$) physics in the two-nucleon system. This physics appears as a string of NN contact interactions of increasing mass dimension, each of which has an LEC associated with it. We then identify the NN potential $V$ at order $D$ as the sum of all 2PI $NN$ graphs---loops plus contacts---up to that order. (Technically 
 an extra step is desirable here, with the 2PI amplitude transformed into an energy-independent potential using e.g. the Okubo formalism~\cite{Ep98}.)
 
 The full NN amplitude is then reconstructed from its 2PI parts via:
\begin{equation}
T({\bf p}',{\bf p};E)=V({\bf p}',{\bf p}) + \int \frac{d^3p''}{(2 \pi)^3} V({\bf p}',{\bf p}'') \frac{1}{E^+ - {\bf p''}^2/M} T({\bf p}'',{\bf p};E),
\label{eq:LSE}
\end{equation}
where an integral over $p_0$ has been performed to combine the two single-nucleon propagators into the propagator of an NN state of energy $E$. Since Eq.~(\ref{eq:LSE}) is just the Lippmann-Schwinger equation, which is equivalent to the Schr\"odinger equation, we now have a non-relativistic quantum mechanics with the NN potential $V$.  The resulting approach has been dubbed ``chiral effective theory" ($\chi$ET) since it is a quantum mechanics at fixed particle number, rather than a quantum field theory. Within $\chi$ET not only $V$, but also quantum-mechanical operators for the interaction of external probes (pions, electrons, photons, \ldots) can be computed up to a given chiral dimension $D$. If this is done to the same order for current operators and for $V$ then the relevant Ward identities are automatically satisfied. But, more generally, this $\chi$PT organization of quantum-mechanical operators is a systematically improvable way to compute few-nucleon structure and reactions at energies of order $m_\pi$. 
 
\begin{figure}[htb]
\begin{center}
\includegraphics[width=12cm]{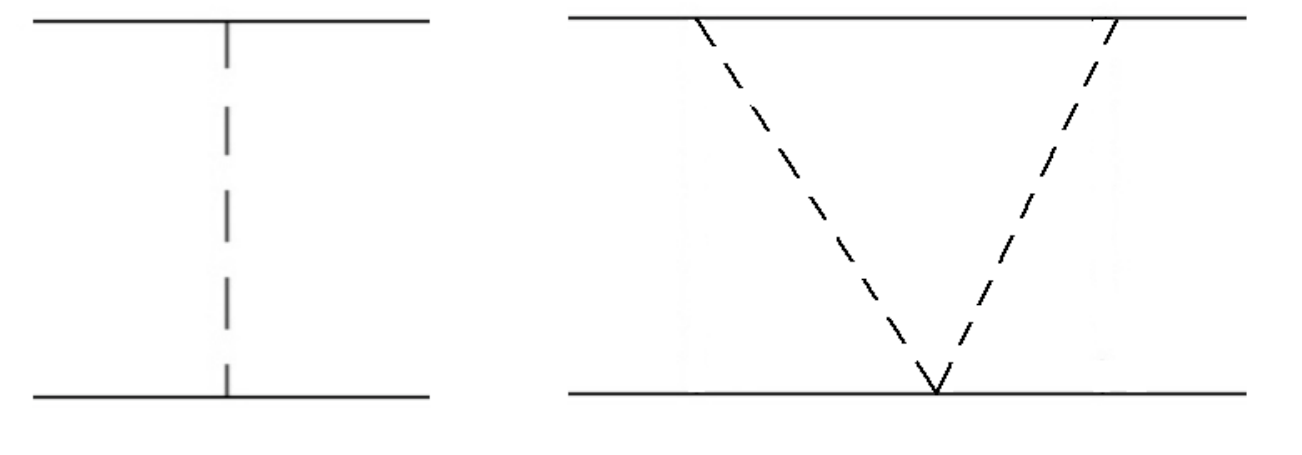}
\end{center}
\caption{\label{fig-NNdiags} Two diagrams that contribute to the long-range part of the NN potential $V$. The diagram on the left is $O(P^0)$, and generates the usual one-pion exchange potential, which occurs in the leading-order piece of $V$. The one on the right is $O(P^2)$, and forms part of the NLO correction to the potential.}
\end{figure}

In Ref.~\cite{Ka97} the potential $V$ was computed at the one-loop level using ${\cal L}_{\pi N}^{(1)}$ and up to one insertion from ${\cal L}_{\pi N}^{(2)}$---reproducing the earlier result of Ref.~\cite{Or96}. The formula (\ref{eq:chiptpc}) tells us that when this is done we have the complete result for $V$ up to $O(P^3)$. Representative diagrams at $O(P^0)$ and $O(P^2)$ are shown in Fig.~\ref{fig-NNdiags}. In Ref.~\cite{Ka97} the potential was then used (in Born approximation) to obtain phase shifts in the higher partial waves of NN scattering. Because of the centrifugal barrier, these partial waves are sensitive mainly to the long-distance part of the force. A good description of a number of partial waves with $l \geq 2$ was found, with the relative contributions of the different chiral orders roughly in agreement with predictions of the chiral expansion.  This suggests that $\chi$PT provides a useful organizing principle for the $r \sim 1/m_\pi$ part of the NN potential. 

This is a key finding, since it opens the way for a treatment of nuclear physics with quantifiable uncertainties. If phase shifts can be obtained up to a fixed order $n$ in the $\chi$PT expansion parameter $P$, then that suggests that the remaining, omitted physics can impact the phase shift only at a level $\sim P^{n+1}$. But the crucial question of how to deal with low partial waves, where the NN interaction is non-perturbative, and in principle the nucleons probe arbitrarily short distances, was not tackled in Ref.~\cite{Ka97}.

In the seminal paper of Ord\'o\~nez {\it et al.}~\cite{Or96} the NN potential was computed up to $O(P^3)$ and Eq.~(\ref{eq:LSE}) solved for phase shifts in a number of NN partial waves. (See also the improved $O(P^3)$ calculation of Ref.~\cite{Ep99}.)  Such a non-perturbative treatment of potentials derived from $\chi$PT has also been advocated in Ref.~\cite{Le97,Ge99,Ol03,Dj07}. It has the advantage that the long-distance parts of $V$, and hence of $T$, are consistent with the pattern of chiral-symmetry breaking in QCD. It also means that for $r \sim 1/m_\pi$ ($\equiv |{\bf p}| \sim m_\pi$) the $\chi$PT counting provides a hierarchy of mechanisms within $V$: the leading-order one-pion exchange is more important than the $O(P^2)$ two-pion exchange which in turn is more important than three-pion exchange, and so forth. 

In practice, in order to solve Eq.~(\ref{eq:LSE}) with the potentials derived from $\chi$PT one must introduce a cutoff, $\Lambda$, on the intermediate states in the integral equation, because  the $\chi$PT potentials grow with the momenta ${\bf p}$ and ${\bf p}'$. The contact interactions in ${\cal L}_{NN}$ should then absorb the dependence of the effective theory's predictions on $\Lambda$ in all low-energy observables, or otherwise we conclude that $\chi$ET is unable to give reliable predictions. In Refs.~\cite{Or96,Ep99,EM02,EM03,Ep05} $V$ was computed to a fixed order, and then the NN LECs that appear in $V$ were fitted to NN data for a range of cutoffs between 500 and 800 MeV. The resulting predictions---especially the ones obtained with the $O(P^4)$ potential derived in Refs.~\cite{EM03,Ep05}---contain very little residual cutoff dependence in this range of $\Lambda$'s, and describe NN data with considerable accuracy. 

However, several recent papers showed that $\chi$ET---at least as presently formulated---does not yield stable predictions once cutoffs larger than $m_\rho$ ($\sim 800$ MeV) are considered~\cite{Towards,ES03,NTvK05,Bi06,PVRA06A,PVRA06B,Ya08,Ya09}. They argue that  the theory described above is not properly renormalized, i.e. the impact of short-distance physics on the results is not under control.
Thus the $\chi$ET described here cannot be used over a wide range of cutoffs. Meanwhile, in Ref.~\cite{EM06} it was argued that the cutoff $\Lambda$ should be varied only in the vicinity of $m_\rho$ in order to obtain an indication of the impact of omitted short-distance physics on the theory's predictions. Since the short-distance physics of the effective theory for $p \gg m_\rho$ is (presumably) completely different to the short-distance physics of QCD itself, considering larger cutoffs that allow momentum integrals to probe $p > m_\rho$ does not yield any additional information regarding the true impact of short-distance physics on observables. Furthermore, Ref.~\cite{EM06} followed Ref.~\cite{Le97} in arguing that working with momenta in the LSE which are demonstrably within the domain of validity of $\chi$PT has the advantage that the $\chi$PT counting must then apply for all parts of $V$: including the contact operators. This justifies $\chi$ET as a systematic theory, but at the cost of limiting its use to cutoffs $m_\pi \ll \Lambda \ll \Lambda_{\chi {\rm SB}}$. Calculations of NN scattering using $\Lambda$'s in this window have been performed in Refs.~\cite{Or96,Ep99,EM02,EM03,Ep05} with, as noted above, considerable phenomenological success. 

Consistent three-nucleon and four-nucleon forces have also been derived in $\chi$ET~\cite{vK94,Ep02,Ro06}. The three-nucleon force enters first at $O(P^3)$, and the four-nucleon force at $O(P^4)$. This provides an EFT-based explanation for the observed hierarchy of nuclear forces: two-nucleon forces provide the bulk of nuclear properties, with three-nucleon (3N) forces accounting for small but crucial corrections, and four-nucleon (4N) forces almost negligible. 3N and 4N forces have been implemented in calculations of the ``low-cutoff" type described in the previous paragraph. For instance, 
the 3N force of Ref.~\cite{Ep02}, in conjunction with the two-nucleon force of Ref.~\cite{EM03}, yields a good description of nuclides up to $A=13$~\cite{NCSM}. An $O(P^3)$ calculation also explains a variety of neutron-deuteron scattering data for neutron beam energies up to about 100 MeV~\cite{Ep02}. Similar success have been achieved for electron-deuteron scattering~\cite{MW01,Ph03,Ph06}, Helium-4 photoabsorption~\cite{QN07}, neutrino-deuteron scattering and breakup~\cite{Parketal}, pion-deuteron scattering at threshold~\cite{We92,pidop3,Be03}, pion photoproduction~\cite{Be95,Be97}, and---as will be discussed extensively in Section~\ref{sec-compton}---coherent Compton scattering from the NN system.
 From all of this evidence we conclude that structure and reactions in few-nucleon systems can be well understood using $\chi$ET---provided the cutoff is kept below about 800 MeV. 

\section{Neutron electromagnetic form factors}

\label{sec-ffs}

The neutron, as its name implies, is an object with zero nett charge. This does not, though, preclude the existence of charged substructure within the neutron. The distribution---and to some extent, motion---of charges within the neutron is encoded in its form factors. Since the neutron is a spin-half particle there are two of these: electric and magnetic in a non-relativistic formulation. These form factors carry the imprint of the chiral dynamics of QCD through its impact on the neutron's internal structure. 

\begin{figure}[htb]
\vskip -0.5cm
\begin{center}
\includegraphics[width=6cm]{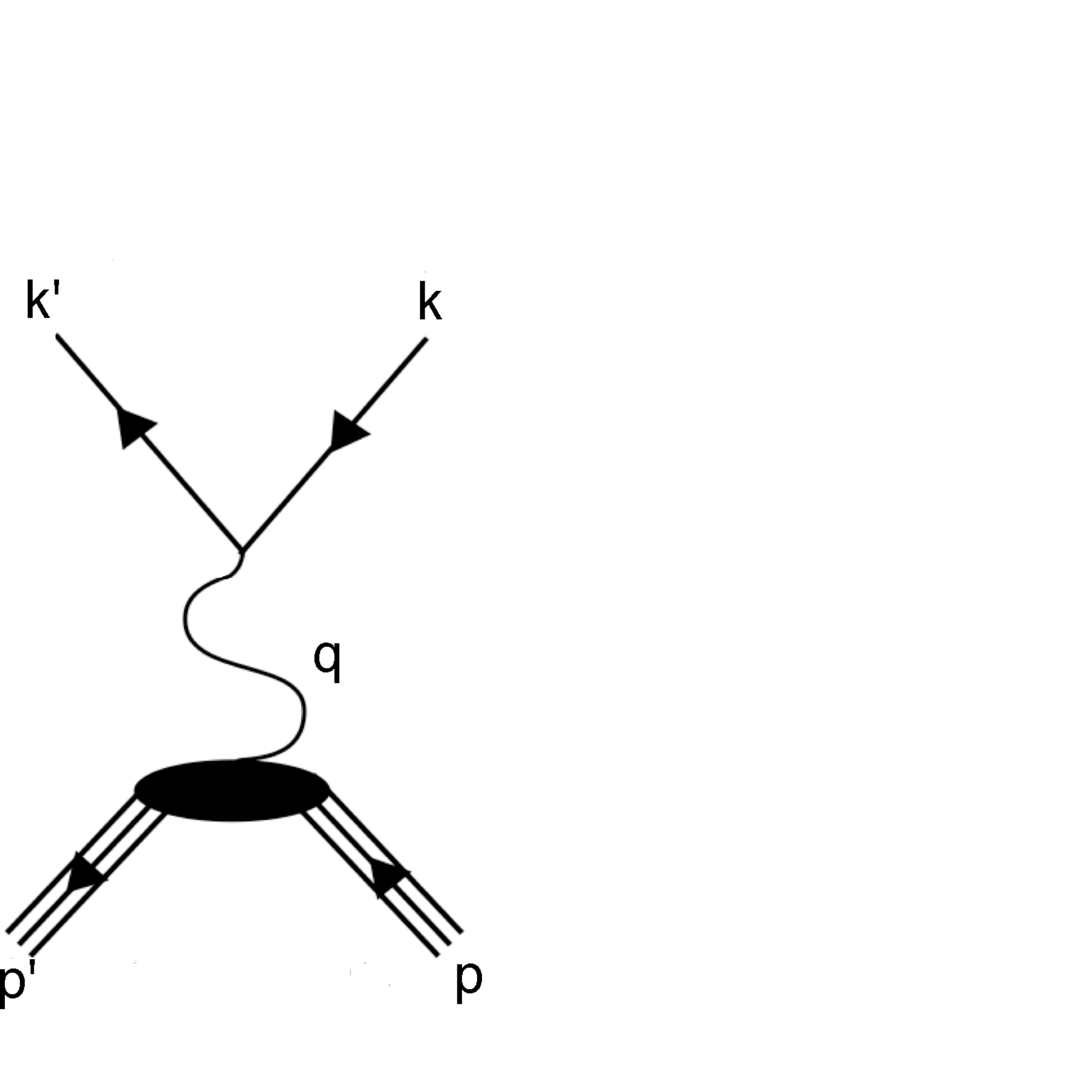}
\end{center}
\vskip -0.7cm
\caption{\label{fig-eNkin} Scattering kinematics for electron-nucleon scattering, in the one-photon exchange approximation. In this diagram the nucleon is denoted by the triple line, and the electron by the single line at the top of the figure. The blob represents the photon-nucleon interaction, $i e J^\mu$.}
\end{figure}

The chiral structure of the neutron would be revealed in electron scattering from neutron targets---if such experiments were feasible. The {\it gedankenscattering} reaction is pictured in Fig.~\ref{fig-eNkin}, where the one-photon-exchange approximation has been assumed. The amplitude is then written as 
\begin{equation}
{\cal M}=-\frac{e^2}{q^2} j^{lep}_\mu J^\mu.
\end{equation}
where $q=k-k'$ is the four-momentum of the exchanged photon (note $q^2 < 0$ for electron scattering) and $j^{lep}_\mu$ is the leptonic current:
\begin{equation}
j_\mu^{lep}=\bar{u}({\bf k}') \gamma_\mu u({\bf k}).
\end{equation}
The most general form of the neutron relativistic current is the usual:
\begin{equation}
J_\mu^n=\bar{u}({\bf p}') \left[F_1^n(q^2) \gamma_\mu + \frac{i}{2M} F_2^n(q^2) \sigma_{\mu \nu} q^\nu\right] u({\bf p}).
\end{equation}
While in the non-relativistic framework being employed here we have:
\begin{equation}
J_\mu^n=\xi_{m_s'}^\dagger \left\{ G_E^n(q^2) v_\mu + \frac{1}{M} G_M^n(q^2) [S_\mu,S_\nu] q^\nu \right\} \xi_{m_s},
\end{equation}
where $\xi$ is, in practice, a Pauli spinor~\footnote{It is a four-component spinor whose lower comonents are zero for the standard choice $v=(1,{\bf 0})$}, and $G_E^n$ and $G_M^n$ are the neutron electric and magnetic form factors.

The following discussion is based on the HB$\chi$PT analysis of Refs.~\cite{Be92,Be98}. For a parallel discussion  within the context of dispersion relations and models on the role of the nucleon's pion cloud in determining electromagnetic form factors see Refs.~\cite{FW03,Ha03}.

\begin{figure}[htb]
\begin{center}
\includegraphics[width=12cm]{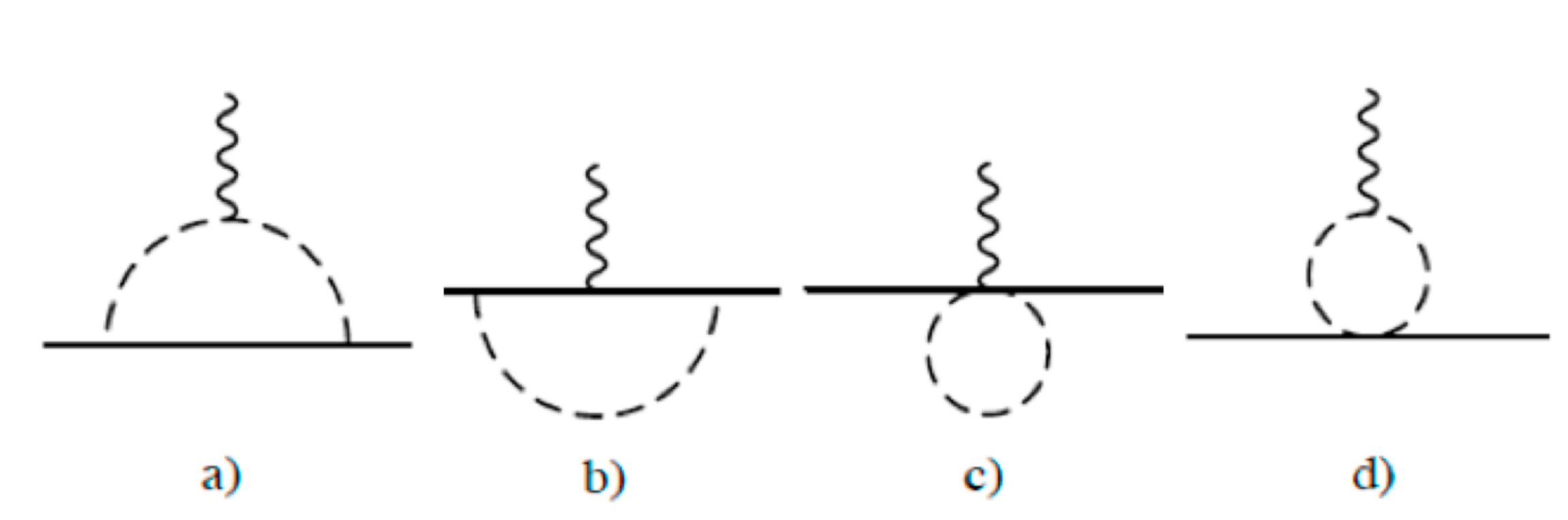}
\end{center}
\caption{\label{fig-oneloopff} Loop diagrams to be computed to obtain the $O(P^3)$ result for the single-nucleon form factors in HB$\chi$PT. All vertices are from ${\cal L}_{\pi N}^{(1)}$ and ${\cal L}_{\pi \pi}^{(2)}$. Picture adapted from Ref.~\cite{Be98}. Used with permission of the authors and Elsevier.}
\end{figure}

For our purposes it is most straightforward to begin with the isoscalar and isovector form factors. These are related to the proton and neutron form factors by:
\begin{eqnarray}
G_{E,M}^s&=&G_{E,M}^p + G_{E,M}^n,\\
G_{E,M}^v&=&G_{E,M}^p - G_{E,M}^n.
\end{eqnarray}
This decomposition is useful because the leading pion-loop effects occur only in the isovector form factors. The quantum numbers of the pion guarantee that the first contribution of the pion cloud to the isoscalar electromagnetic form factors occurs only at the two-loop level. The one-loop calculation was first carried out in Ref.~\cite{Be92}, and was extended and refined in Ref.~\cite{Be98}. Combining the $O(P^3)$ diagrams shown in Fig.~\ref{fig-oneloopff} with the relevant tree graphs  yields the expression
\begin{eqnarray}
G_E^v=1 + \frac{q^2}{(4 \pi f_\pi)^2} \left\{\kappa_v \left(\frac{2 \pi f_\pi}{M}\right)^2 - \frac{2}{3} g_A^2 - 2 B_{10}^{(r)}(\mu) - \left[\frac{5}{3}g_A^2 + \frac{1}{3} \right] \log \left(\frac{m_\pi}{\mu}\right)\right\} \nonumber\\
+ \frac{1}{(4 \pi f_\pi)^2} \int_0^1 dx \, \left[(3 g_A^2 + 1) m_\pi^2 - q^2 x (1-x) (5 g_A^2 + 1)\right] \log \left[\frac{m_\pi^2 - q^2 x (1-x)}{m_\pi^2}\right]\nonumber\\
\label{eq:GEv}
\end{eqnarray}
Here we have worked in the Breit frame, and so $q^2=-|{\bf q}|^2 \equiv Q^2$, with ${\bf q}={\bf p}' - {\bf p}$ the three-momentum transfer to the struck neutron. 

The first term, ``1" is guaranteed by charge conservation. The second piece gives the isovector charge radius, according to the usual definition:
\begin{equation}
G_E^v=1 - \frac{1}{6} \langle (r_E^v)^2 \rangle q^2 + O(q^4).
\end{equation}
The short-distance contribution to the isovector charge radius, $B_{10}^{(r)}$, renormalizes the divergent loop integrals. It arises from a term in ${\cal L}_{\pi N}^{(3)}$ of the form
\begin{equation}
{\cal L}_{\pi N}^{(3)} \sim B_{10} \bar{H} D^\alpha f_{\alpha \nu}^+ v^\nu H.
\label{eq:B10}
\end{equation}

The formula (\ref{eq:GEv}) is obtained using dimensional regularization with the modified minimal subtraction of $\chi$PT for the calculation of the loop integrals. The physical result for $G_E^v$ is not dependent on the choice of regularization and renormalization scheme, but the finite parts of $B_{10}^{(r)}(\mu)$ will in general be so. The $\mu$ dependence is not, however, affected by the particular variant of minimal subtraction employed, since the $\mu$ dependence of $B_{10}^{(r)}(\mu)$ must cancel the $\mu$ dependence of the logarithimc term in the curly brackets that contributes to the isovector charge radius.

Note that $B_{10}$ only affects the charge radius, and all higher moments of the charge distribution are given---at $O(P^3)$ accuracy---by the physics of the pion-loop graphs shown in Fig.~\ref{fig-oneloopff}.
The presence of this short-distance contribution to $\langle (r_E^v)^2 \rangle$ means that the isovector charge radius cannot be predicted in $\chi$PT. In fact, the corresponding calculation for the isoscalar form factor gives purely short-distance effects:
\begin{equation}
G_E^s=1 - \frac{q^2}{(4 \pi f_\pi)^2} \left[\kappa_s \left(\frac{2 \pi f_\pi}{M}\right)^2 - 4 \tilde{B_1}\right].
\end{equation}
Here $\tilde{B}_1$ is the coefficient of the $\gamma NN$ contact interaction that is the isoscalar analog of Eq.~(\ref{eq:B10}). However, since there is no loop effect at this order $\tilde{B}_1$ is a finite LEC.

Thus the prediction for the neutron electric form factor, up to $O(P^3)$,  in HB$\chi$PT  is:
\begin{eqnarray}
G_E^n=-\frac{q^2}{(4 \pi f_\pi)^2} \tilde{B}_e^n - \frac{q^2 \kappa^n}{4 M^2}
 - \frac{1}{2 (4 \pi f_\pi)^2} \int_0^1 dx \, \left[(3 g_A^2 + 1) m_\pi^2 \right.\nonumber\\
\qquad  \qquad \qquad \quad \left. - q^2 x (1-x) (5 g_A^2 + 1)\right]
 \log \left[\frac{m_\pi^2 - q^2 x (1-x)}{m_\pi^2}\right],
 \label{eq:GEn}
\end{eqnarray}
with $\tilde{B}_e^n$ a cobination of $\tilde{B}_1$, $\bar{B}_{10}$, and finite terms with fixed coefficients. The measured value of the neutron charge radius squared is $\langle (r_E^n)^2 \rangle=-0.115(4)~{\rm fm}^2$~\cite{Kopecky,BHM}. Comparing this with the result of Eq.~(\ref{eq:GEn}) reveals that, numerically, the second term on the the first line of Eq.~(\ref{eq:GEn})---the Foldy term---dominates the overall result, and
$\tilde{B}_e^n$ is unnaturally small. The non-analytic behaviour beyond $O(q^2)$ encoded in the integral in Eq.~(\ref{eq:GEn}) is a prediction of $\chi$PT at this order, and arises because of the spontaneously broken chiral symmetry in QCD. 

In the case of the magnetic form factor the only counterterm necessary is the one already written above in ${\cal L}_{\pi N}^{(2)}$. It renormalizes the anomalous magnetic moment, i.e. the value of the form factor at $q^2=0$. Consequently, for the neutron, we find:
\begin{equation}
G_M^n=\kappa^n + g_A^2\frac{2 \pi M}{(4 \pi f_\pi)^2} \int_0^1 dx \, \left(\sqrt{m_\pi^2 - q^2 x(1-x)} - m_\pi \right).
\label{eq:GMn}
\end{equation}
Expanding in powers of $q^2$ we find that the magnetic radius of the neutron blows up in the chiral limit as $m_q^{-1/2}$. This is to be compared to the $\log(m_q)$ dependence of the electric radius  seen in Eq.~(\ref{eq:GEv}). These chiral-limit divergences of observables---together with the associated pre-factors---are model-independent predictions of $\chi$PT, since they result from the long-distance pieces of the relevant quantum loops. 
 
In Eq.~(\ref{eq:GMn}) short-distance physics contributes only to $\kappa^n$. All other moments of the magnetization distribution are dominated by pion-cloud physics, and have only small ($O(P^4)$ and beyond) contributions from other baryons and mesons. Of course, this statement applies only to the regime $|{\bf q}| \sim m_\pi$. Once $|{\bf q}| \approx 600$ MeV a variety of other dynamics needs to be taken into consideration, since at that momentum scale the electromagnetic probe can excite shorter-degrees of freedom in the neutron's electromagnetic structure. 

\begin{figure}[htb]
\begin{center}
\includegraphics[width=15cm]{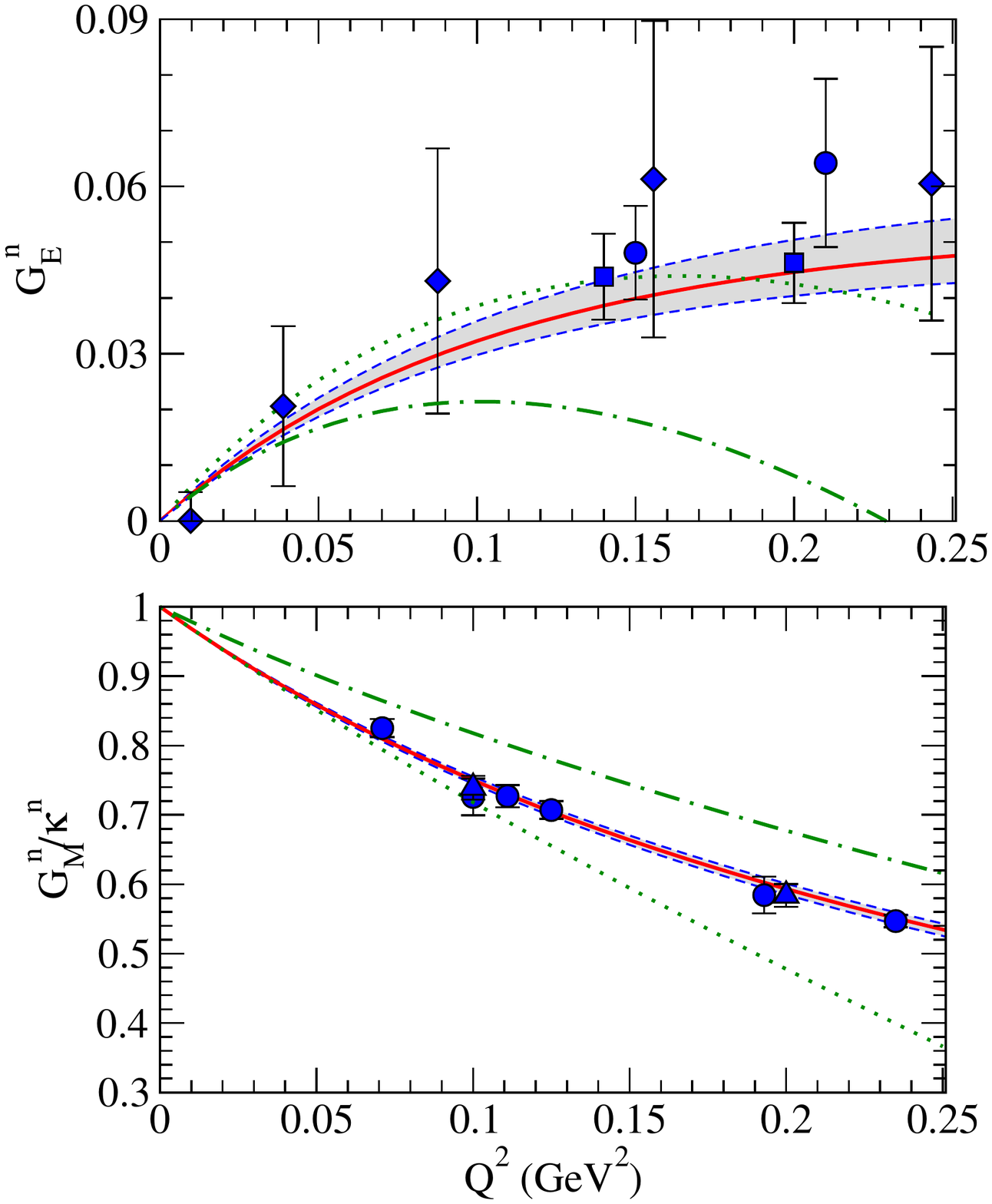}
\end{center}
\caption{\label{fig-neutronffs} Results for $G_E^n$ (upper panel) and $G_M^n$ (lower panel) obtained in HB$\chi$PT up to $O(P^3)$ (green dot-dashed line) as compared to the dispersion-relation fit of Belushkin, Hammer, and Mei\ss ner (red line, with error bands represented by the shaded region)~\cite{BHM}. The (green) dotted line in the upper panel corresponds to an $O(P^3)$ calculation, but with an input value of $\langle (r_E^n)^2 \rangle=-0.155~{\rm fm}^2$. The $G_E^n$ measurements represented by the circles are from Refs.~\cite{He99,Pa99}, while the squares are the recent BLAST results~\cite{Ge08}. The diamonds are from the analysis of data on the deuteron quadrupole form factor of Ref.~\cite{SS01}. In the lower panel the data included in the fit of Ref.~\cite{BHM} is shown. The circles are results from deuteron breakup measurements (for details see Ref.~\cite{FW03}). The triangles are from the ${}^3$He$(e,e')$ measurement of Ref.~\cite{Xu}.
The dotted line in the lower panel is a partial $O(P^4)$ HB$\chi$PT calculation of $G_M^n$ in which a $\gamma^*$nn contact interaction is included so as to bring the result for the neutron's magnetic radius into agreement with the central value obtained in Ref.~\cite{BHM}. }
\end{figure}

The $O(P^3)$ HB$\chi$PT predictions for $G_E^n$ and $G_M^n$ are plotted in Fig.~\ref{fig-neutronffs}, together with some ``data" for both---on which, more below. Also shown is a recent dispersion-relation fit to a large database of electron-scattering experiments~\cite{BHM}. 

The upper panel of Fig.~\ref{fig-neutronffs} shows that the $O(P^3)$ result for $G_E^n$ has the same feature as this dispersion-relation description, namely a rise at low-$Q^2$, followed by a turn over. But the HB$\chi$PT prediction turns over too quickly. However, adjusting the input neutron charge radius to a slightly larger value produces a marked effect. And since $\langle (r_E^n)^2 \rangle$ results from a sizeable cancellation between isovector and isoscalar pieces of short-distance physics, this finding suggests that the neutron electric form-factor data can be accommodated within a ``natural" scenario for that higher-energy dynamics. Indeed, the difficulty of describing $G_E^n$ accurately within HB$\chi$PT stems partly from these cancellations, which lead to a $G_E^n$ that is small throughout the entire realm of $\chi$PT's applicability. This means that higher-order contributions will have a disproportionately large impact on the final answer. 

The neutron magnetic form factor, which is order one in this domain, is more fertile ground for $\chi$PT. And indeed, the $O(P^3)$ prediction for $G_M^n$ has the right general shape, but yields a magnetic radius of 
\begin{equation}
\langle (r_M^n)^2 \rangle \equiv -\frac{6}{\kappa^n} \frac{d G_M^n}{dQ^2}=-\frac{g_A^2 M}{16 f_\pi^2 \kappa_n \pi m_\pi}=0.51~{\rm fm}^{2},
\end{equation}
which is a little smaller than the magnetic radius obtained from the dispersion-relation analysis. However, it must be remembered that there is an $O(P^4)$ short-distance contribution to this quantity. Including that contribution in the calculation of $G_M^n$ yields the dotted curve shown in Fig.~\ref{fig-neutronffs}. This can be considered a partial $O(P^4)$ calculation of $G_M^n$, since it does not include the additional loop mechanisms that are present at this order. But even this partial $O(P^4)$ computation shows that the shift due to short-distance effects is about 20\% at $Q^2=0.1$ GeV$^2$---a magnitude that is consonant with this being a contribution of $O(Q^2/\Lambda_{\chi {\rm SB}}^2)$ to $G_M^n/\kappa_n$. In fact, the $O(P^4)$ contact interaction that shifts the magnetic radius to the empirical value is associated with a momentum scale of order 1 GeV. The LEC that is the coefficient of this contact interaction is certainly natural with respect to $\Lambda_{\chi {\rm SB}}$.  But by $Q^2 \approx 0.25$ GeV$^2$ the $O(P^4)$ contribution is as large as the $O(P^3)$ result---a clear sign that the $\chi$PT expansion is breaking down. The results for $G_M^n$ therefore confirm the $\chi$PT picture that the dominant effect in the neutron's magnetization distribution is the pion cloud, while also showing that this picture is limited to momentum transfers below 500 MeV.
 
The results reported here are rigorous within HB$\chi$PT. Ref.~\cite{Be98} showed that the inclusion of Delta(1232) degrees of freedom in the effective field theory does not improve the agreement with data---the short-distance physics in these observables is associated with higher-mass states. The impact of corrections associated with nucleon recoil on neutron form-factor predictions is discussed in Refs.~\cite{KM01,Sc05}. These corrections are appreciable in the case of $G_E^n$, and there the $q/M$ series can be reorganized into a form that improves the agreement of the $\chi$PT predictions with data. The tree-level contribution of vector mesons is also considered in Refs.~\cite{KM01,Sc05} and it assists in the description of $G_M^n$.

The ``data" points represented by circles and squares in Fig.~\ref{fig-neutronffs} are obtained via electron-induced breakup reactions on deuterium targets. It needs to be emphasized that $\chi$PT calculation of these reactions would allow a completely consistent treatment of the pion dynamics that generates the higher-order terms in the neutron form factor and the pion-exchange currents which play, e.g. an important role in $\gamma d \rightarrow n p$ at threshold~\cite{Pa94}. These exchange currents occur at $O(P^3)$ in the irreducible kernel for the interaction of real and virtual photons with the NN system, and so have the same chiral order as the single-nucleon loop effects that have been highlighted in this Section.
But, no comprehensive analysis of $e d \rightarrow e' np$ has been performed within $\chi$ET. Such an analysis, and its extension to the three-body system, would allow a consistent treatment of the chiral structure of the neutron and the target nucleus. It would also facilitate reliable estimates of the theoretical uncertainties in the neutron form-factor extraction.

The diamonds in the upper panel of Fig.~\ref{fig-neutronffs} are obtained from an analysis of the deuteron's quadrupole form factor, $G_Q$~\cite{SS01}. In this case the AV18 potential~\cite{AV18} was used to generate the deuteron's wave function, and the important correction to the deuteron's charge operator~\cite{Ph03,Ri84,Ad93} was included in the calculation. However, this calculation fails to reproduce the deuteron's quadrupole moment, and this could have some consequences for the $Q^2$-dependence of the deuteron $G_Q$~\cite{Ph06}. $\chi$ET describes data on the ratio of deuteron charge to quadrupole form factors $G_C/G_Q$ well up to momentum transfers of about 0.3 GeV$^2$~\cite{Ph06}. A $\chi$ET extraction of $G_E^{(s)}$ from $G_Q$, and possibly from the $A(Q^2)$ data of Ref.~\cite{Pl90} and subsequent studies, could perhaps provide useful information on $G_E^{n}$ in this low-$Q^2$ regime.

\section{Compton scattering from the neutron}

\label{sec-compton}

The neutron has zero charge, but its composite structure means that it has a non-zero Compton scattering amplitude. The standard decomposition for the neutron Compton amplitude involves six non-relativistic invariants:
\begin{eqnarray}
&&T_{\gamma N}= A_1 {\bf \epsilon}' \cdot {\bf \epsilon}
            +A_2{\bf \epsilon} \cdot \hat{k} \, {\bf \epsilon} \cdot \hat{k'}
             +iA_3 {\bf \sigma} \cdot ({\bf \epsilon}' \times {\bf \epsilon})
             +iA_4{\bf \sigma} \cdot (\hat{k'}\times\hat{k})\, {\bf \epsilon'} \cdot {\bf \epsilon} 
    \nonumber \\
 & &
\!\!\!\!\!\!\!\!\!
     +iA_5 {\bf \sigma} \cdot [({\bf \epsilon}' \times\hat{k})\,{\bf \epsilon} \cdot\hat{k'}
            -({\bf \epsilon} \times\hat{k'})\, {\bf \epsilon}' \cdot\hat{k}]
     +iA_6 {\bf \sigma} \cdot [({\bf \epsilon}' \times\hat{k'})\,{\bf \epsilon} \cdot\hat{k'}
                        -({\bf \epsilon} \times\hat{k})\, {\bf \epsilon}' \cdot\hat{k}],\nonumber\\
\label{eq:Ti}
\end{eqnarray}
where ${\bf k}$ and ${\bf \epsilon}$ (${\bf k}'$ and ${\bf \epsilon}'$) are the three-momentum and polarization vector of the incoming (outgoing) photon. The $A_i$'s, $i=1\ldots 6$, are scalar functions of photon energy and scattering
angle.

\begin{figure}[htbp]
\begin{center}
\includegraphics[width=10cm]{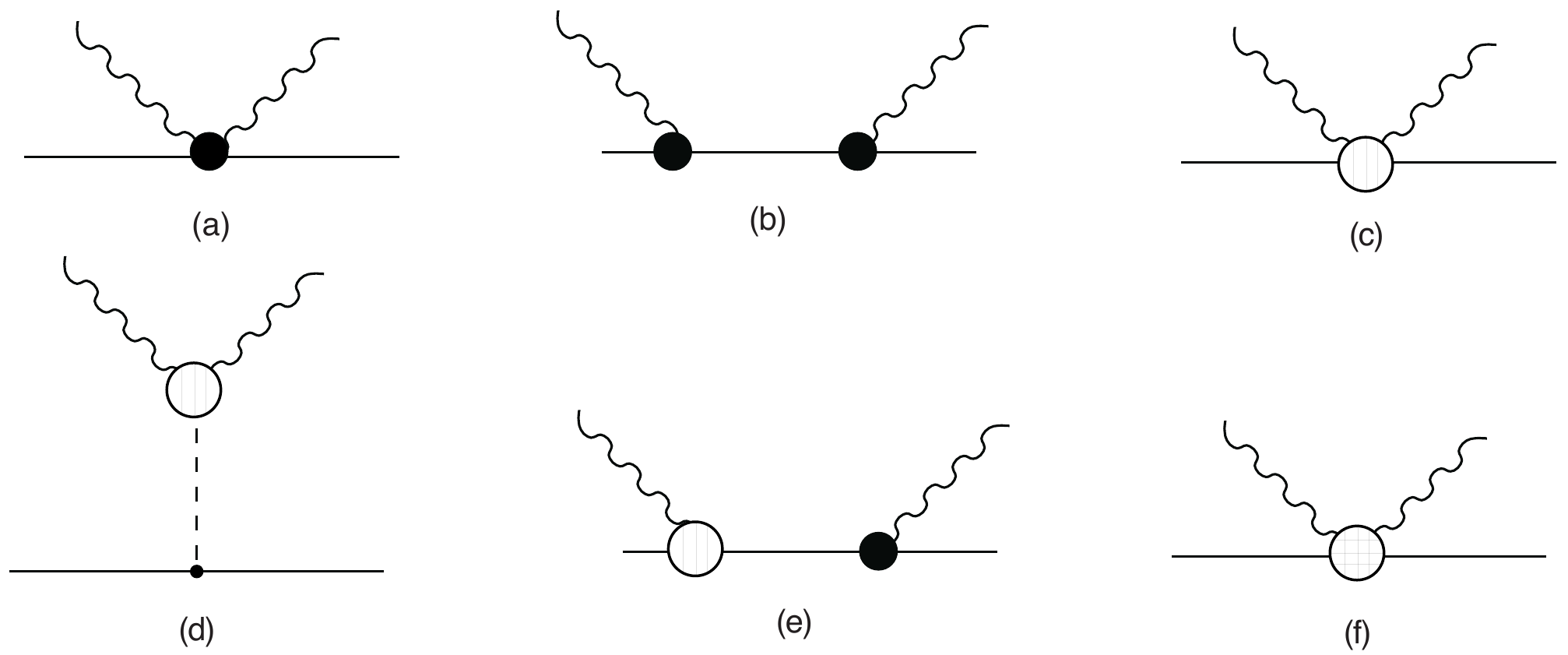}
\end{center}
\caption{\label{fig-protborn} Tree diagrams that contribute to Compton
scattering in the $\epsilon\cdot v=0$ gauge up to $O(P^4)$. Small dots
are vertices from ${\cal L}_{\pi N}^{(1)}$, larger dots are vertices from
${\cal L}_{\pi N}^{(2)}$, the sliced dots are vertices from ${\cal L}_{\pi N}^{(3)}$
and the hatched dot is a vertex from ${\cal
L}_{\pi N}^{(4)}$. Crossed graphs and graphs which differ only in the ordering
of vertices are included in
the calculation, but are not shown here. Figure from Ref.~\cite{Be04}, reprinted with permission of Elsevier.}
\end{figure}

\begin{figure}[htbp]
\begin{center}
\includegraphics[width=12cm]{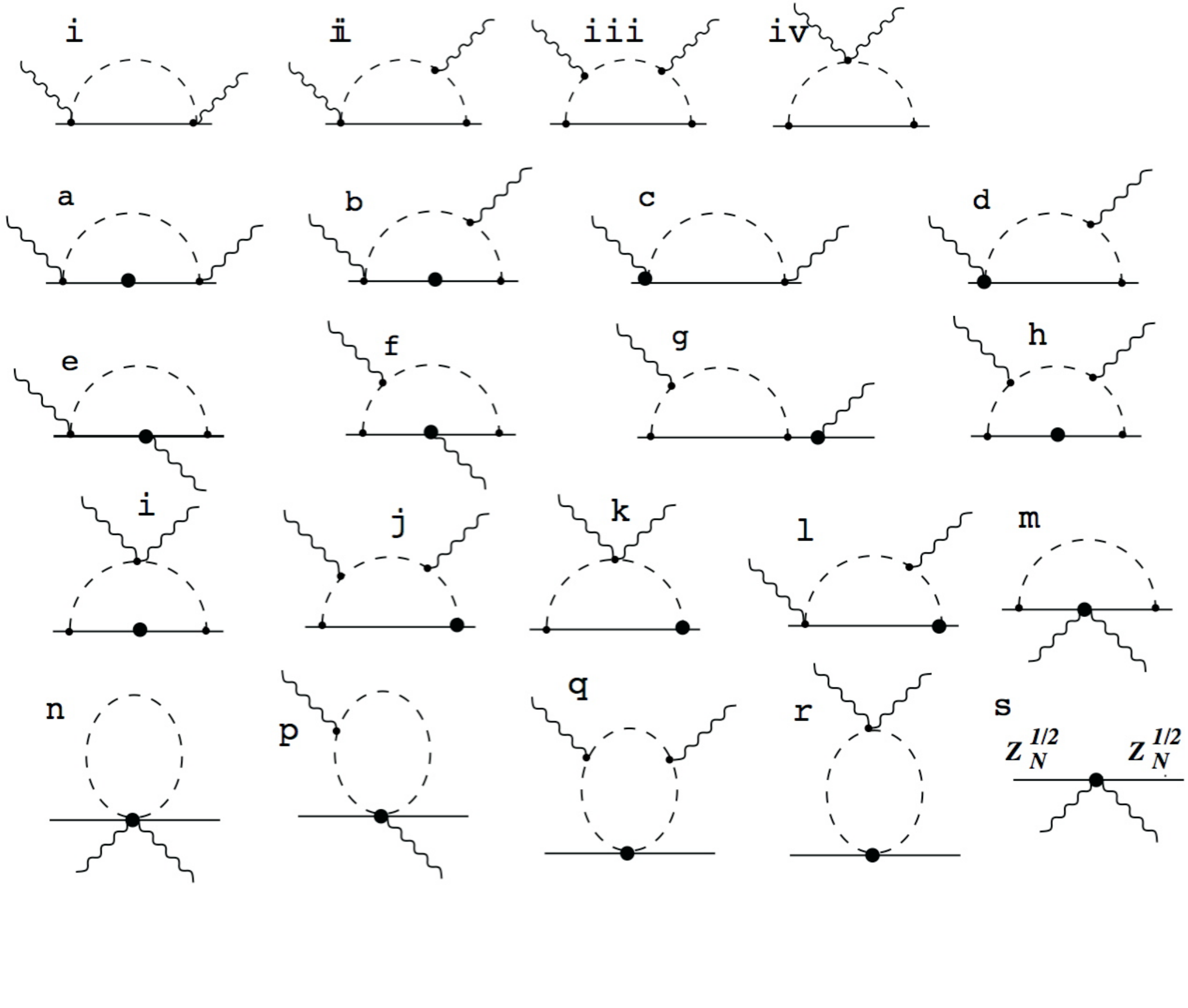}
\end{center}
\caption{\label{fig-protloop} Diagrams which contribute to nucleon Compton
scattering in the $\epsilon\cdot v=0$ gauge at 3rd (i-iv) and 4th
(a-s) order. Vertices are labeled as in Fig.~\ref{fig-protborn}. Figure reprinted from Ref.~\cite{Be04}, with permission from Elsevier.}
\end{figure}

At photon energies $\sim m_\pi$ Compton scattering from the neutron is driven by photon interactions with the neutron's pion cloud. The dominant effects come from the $O(P^3)$ mechanisms depicted in the diagrams (a)--(d) of Fig.~\ref{fig-protborn} and the first line of Fig.~\ref{fig-protloop}. These loop diagrams individually contain divergences, but the divergences cancel in the sum. Indeed, they must do so, because no gauge-invariant $\gamma$n contact interaction can be constructed at $O(P^3)$: at this level of accuracy there is no short-distance contribution to neutron Compton scattering. Consequently, up to $O(P^3)$ there is a $\chi$PT prediction for this process. It, together with the associated proton result, is given by the following (Breit-frame) expressions for $A_1$--$A_6$~\cite{Be93,McG01}:
\begin{eqnarray}
A_1 &=& -\frac{{\mathcal Z}^2 e^2}{M}
        +\frac{g_A^2m_\pi e^2}{8\pi f_\pi^2}
         \left\{ 1- \sqrt{1-\Upsilon^2}
                 +\frac{2-t}{\sqrt{-t}}
                  \left[\frac{1}{2} \arctan \frac{\sqrt{-t}}{2}
                        -I_1(\Upsilon,t) \right]\right\}, 
\nonumber \\
A_2 &=&
        -\frac{e^2 g_A^2 \omega^2}{8\pi f_\pi^2 m_\pi}
         \frac{2-t}{(-t)^{3/2}}
         \left[I_1(\Upsilon,t)- I_2(\Upsilon,t)\right],
\nonumber \\
A_3 &=& \frac{e^2 \omega}{2M^2} [{\mathcal Z}({\mathcal Z}+2\kappa
)-({\mathcal Z}+\kappa)^2 \cos \theta]
        +\frac{{(2{\mathcal Z} -1)}e^2 g_A m_\pi}{8\pi^2 f_\pi^2} \frac{\Upsilon t}{1-t}
\nonumber \\
    & &  +\frac{e^2 g_A^2 m_\pi}{8\pi^2 f_\pi^2}
         \left[ \frac{1}{\Upsilon} \arcsin^2\Upsilon- \Upsilon +2\Upsilon^4
\sin^2 \theta I_3(\Upsilon,t)\right],
\nonumber \\
A_4 &=& -\frac{({\mathcal Z}+\kappa )^2 e^2 \omega}{2M^2}
        +\frac{e^2 g_A^2 \omega^2}{4\pi^2 f_\pi^2m_\pi} I_4(\Upsilon,t),
\nonumber \\
A_5 &=& \frac{({\mathcal Z}+\kappa )^2 e^2 \omega}{2M^2}
        -\frac{{(2{\mathcal Z} -1)}e^2 g_A \omega^2}{8\pi^2 f_\pi^2 m_\pi} \frac{\Upsilon}
        {(1-t)}\nonumber\\
        &&-\frac{e^2 g_A^2 \omega^2}{8\pi^2 f_\pi^2m_\pi}
          [I_5(\Upsilon,t)-2\Upsilon^2\cos\theta I_3(\Upsilon,t)],
\nonumber \\
A_6 &=& -\frac{{\mathcal Z}({\mathcal Z}+\kappa ) e^2 \omega}{2M^2}
        +\frac{{(2{\mathcal Z} -1)}e^2 g_A \omega^2}{8\pi^2 f_\pi^2 m_\pi}\frac{\Upsilon}
         {(1-t)} \nonumber \\
        &&+\frac{e^2 g_A^2 \omega^2}{8\pi^2 f_\pi^2m_\pi}
          [I_5(\Upsilon,t)-2\Upsilon^2 I_3(\Upsilon,t)],
\label{eq:As}
\end{eqnarray}
where $\Upsilon=\omega/m_\pi$, $t=-2 \Upsilon^2 (1 - \cos \theta)$, and 
\begin{eqnarray}
I_1(\Upsilon,t) &=& \int_0^1  dz \,
             \arctan \frac{(1-z)\sqrt{-t}}{2\sqrt{1-\Upsilon^2 z^2}},
\nonumber \\
I_2(\Upsilon,t) &=& \int_0^1  dz \,
             \frac{2(1-z)\sqrt{-t(1-\Upsilon^2z^2)}}{4(1-\Upsilon^2 z^2)-t(1-z)^2},
\nonumber \\
I_3(\Upsilon,t) &=& \int_0^1  dx \, \int_0^1  dz \,
             \frac{x(1-x)z(1-z)^3}{S^3}
             \left[ \arcsin \frac{\Upsilon z}{R}+ \frac{\Upsilon zS}{R^2}\right],
\nonumber \\
I_4(\Upsilon,t) &=& \int_0^1  dx \, \int_0^1  dz \,
             \frac{z(1-z)}{S}\arcsin \frac{\Upsilon z}{R},
\nonumber \\
I_5(\Upsilon,t) &=& \int_0^1  dx \, \int_0^1  dz \,
             \frac{(1-z)^2}{S}\arcsin \frac{\Upsilon z}{R},
\label{eq:Is}
\end{eqnarray}
with
\begin{equation}
S=\sqrt{1-\Upsilon^2 z^2-t(1-z)^2x(1-x)}, \quad
R=\sqrt{1-t(1-z)^2x(1-x)}. \label{eq:sr}
\end{equation}
Here and below formulae are given in Coulomb gauge, for real photons, so the polarization four-vectors and photon four-momenta satisfy: 
\begin{equation}
k^2=0; \qquad k'^2=0; \qquad v \cdot \epsilon=0; \qquad v \cdot \epsilon' =0.
\end{equation}

Note the impact of the Goldstone bosons of QCD---the pions---on the neutron Compton amplitude. The theory predicts that the Compton amplitude contains a cusp at $\omega=m_\pi$. That cusp is a consequence of the opening of the photoproduction channel at that energy~\footnote{In HB$\chi$PT the threshold is not at precisely the right photon energy, but this can be corrected through a resummation of higher-order terms that is motivated by a modified power counting in the vicinity of a threshold~\cite{Be93}.}, but its consequences are felt even for $\omega$ well below $m_\pi$. Thus Compton scattering---alone among reactions with real photons---probes chiral dynamics at photon energies well below the pion threshold. 

The most famous manifestation of this chiral dynamics in neutron Compton scattering occurs in the neutron polarizabilities. The low-energy expansion of the $A_i$'s defines the electric and magnetic polarizabilities $\alpha$ and $\beta$ as well as the four ``spin polarizabilities" $\gamma_1$--$\gamma_4$. The former occur as coefficients of the $\omega^2$ terms in $A_1$ and $A_2$, while the latter are coefficients of the $\omega^3$ terms in $A_3$--$A_6$---once the Born terms are accounted for. For the Breit-frame amplitude
such an $\omega$-expansion gives (keeping terms up to $O(1/M^3)$)~\cite{McG01,Be04}:
\begin{eqnarray}
A_1(\omega,\theta)\!&=&\!-{{\cal Z}^2e^2 \over M}+{e^2\over 4M^3}
\Bigl(({\cal Z}+\kappa)^2(1+\cos\theta)-{\cal Z}^2\Bigr)(1-\cos\theta)\,\omega^2\nonumber\\
&&\qquad \qquad \qquad \qquad + 4\pi(\alpha + \beta \, \cos\theta)\omega^2+
O(\omega^4),\nonumber \\ 
A_2(\omega,\theta)\!&=&\!{e^2 \over 4M^3}\kappa(2{\cal Z}+\kappa) \omega^2 \cos\theta -4\pi\beta \omega^2 \,+
O(\omega^4),\nonumber\\ 
A_3(\omega,\theta)\!&=&\!  {e^2 \omega \over
2M^2}\Bigl({\cal Z}({\cal Z}+2\kappa)-({\cal Z}+\kappa)^2 \cos\theta\Bigr)
+A_3^{\pi^0}\nonumber\\
&& \qquad \qquad \qquad \qquad 
+ 4\pi\omega^3(\gamma_1 - (\gamma_2 + 2 \gamma_4) \, \cos\theta)+
O(\omega^5),\nonumber\\ A_4(\omega,\theta)\!&=&\! -{e^2 \omega \over 2
M^2 }({\cal Z}+\kappa)^2 + A_4^{\pi^0}+4\pi\omega^3 \gamma_2 +
O(\omega^5),\nonumber\\ A_5(\omega,\theta)\!&=&\! {e^2 \omega \over
2 M^2 }({\cal Z}+\kappa)^2 + A_5^{\pi^0}+4\pi\omega^3\gamma_4 +
O(\omega^5),\nonumber\\ A_6(\omega,\theta)\!&=&\! -{e^2 \omega \over
2 M^2 }{\cal Z}({\cal Z}+\kappa) +A_6^{\pi^0}+4\pi\omega^3\gamma_3 +
O(\omega^5),
\label{eq:Asinw}
\end{eqnarray}
with $A_3^{\pi^0}$--$A_6^{\pi^0}$ the contributions from the pion-pole
graph, Fig.~\ref{fig-protborn}(d).

At $O(P^3)$ the $\chi$PT predictions for these polarizabilities are straightforwardly obtained from a Taylor series of Eq.~(\ref{eq:As}) in powers of $\omega/m_\pi$.
The following results for the polarizabilities can be
obtained~\cite{Be91,Be92} by matching Eq.~(\ref{eq:As}) to
Eq.~(\ref{eq:Asinw}):
\begin{eqnarray}
\alpha^p=\alpha^n=\frac{5 e^2 g_A^2}{384 \pi^2 f_\pi^2 m_\pi}
&=&12.2 \times
10^{-4} \, {\rm fm}^3, \label{eq:alphaOQ3} \nonumber \\
\beta^p=\beta^n=\frac{e^2 g_A^2}{768 \pi^2 f_\pi^2 m_\pi}&=& 1.2
\times
10^{-4} \, {\rm fm}^3, \label{eq:betaOQ3} \nonumber \\
\gamma^p_1=\gamma^n_1=\frac{e^2 g_A^2}{98 \pi^3 f_\pi^2 m_\pi^2} &=&4.4
\times
10^{-4} \, {\rm fm}^4, \label{eq:gamma1OQ3} \nonumber \\
\gamma_2^p=\gamma_2^n=\frac{e^2 g_A^2}{192 \pi^3 f_\pi^2 m_\pi^2} &=&2.2
\times
10^{-4} \, {\rm fm}^4, \label{eq:gamma2OQ3} \nonumber \\
\gamma_3^p=\gamma_3^n=\frac{e^2 g_A^2}{384 \pi^3 f_\pi^2 m_\pi^2} &=&1.1
\times
10^{-4} \, {\rm fm}^4, \label{eq:gamma3OQ3} \nonumber \\
\gamma_4^p=\gamma_4^n=-\frac{e^2 g_A^2}{384 \pi^3 f_\pi^2 m_\pi^2}
&=&-1.1 \times 10^{-4} \, {\rm fm}^3. \label{eq:gamma4OQ3}
\end{eqnarray}
As with the charge and magnetic radii electromagnetic polarizabilities diverge in the chiral limit of QCD---with the coefficient of the divergence predicted by HB$\chi$PT. That divergence is particularly dramatic ($\sim 1/m_q$) in the case of the spin polarizabilities, and this could have interesting consequences for lattice simulations of these quantities~\cite{De06}. And
the HB$\chi$PT result for the $\gamma$-nucleon amplitude actually gives much information besides the predictions (\ref{eq:gamma4OQ3}), since HB$\chi$PT predicts the full dependence on the parameter $\omega/m_\pi$. 

But $\alpha^p$ and $\beta^p$ can be extracted from low-photon-energy proton Compton-scattering experiments. Such measurements are particularly revealing because at $O(P^4)$ $\alpha$ and $\beta$ receive their first correction from short-distance physics. They thus result from a fascinating interplay of the pion-cloud dynamics that yields the dominant effects of Eq.~(\ref{eq:alphaOQ3}), and shorter-distance, higher-$\chi$PT-order mechanisms that correct those predictions.

Elsewise at $O(P^4)$, nucleon-pole graphs \ref{fig-protborn}(e) and
the fixed-coefficient piece of the seagull depicted in
Fig.~\ref{fig-protborn}(f) give further terms in the $1/M$-expansion of
the relativistic Born amplitude. There are also a number of
fourth-order one-pion loop graphs (graphs a-r of Fig.~\ref{fig-protloop}). All
of the graphs \ref{fig-protloop}a--\ref{fig-protloop}r involve vertices from
${\cal L}_{\pi N}^{(2)}$. In particular, the expressions for
diagrams \ref{fig-protloop}n--\ref{fig-protloop}r contain the LECs $c_1$,
$c_2$, and $c_3$,
and in diagrams \ref{fig-protloop}e--\ref{fig-protloop}g 
the nucleon
anomalous magnetic moments enter. 

Divergences in graphs \ref{fig-protloop}a--\ref{fig-protloop}f, 
\ref{fig-protloop}h--\ref{fig-protloop}k, \ref{fig-protloop}p, and \ref{fig-protloop}q 
are renormalized by counterterms from ${\cal L}_{\pi N}^{(4)}$.  The
overall result---first derived in Ref.~\cite{Be93}---is
\begin{eqnarray}
&& 4\pi(\alpha+\beta)^{(4)}= \delta \alpha^s(\mu) + \delta \beta^s(\mu)
+ \tau_3 \left(\delta \alpha^v(\mu) 
+ \delta \beta^v(\mu)\right)\nonumber\\
&& +  {e^2\over 16\pi^2f_\pi^2}\left\{{g_A^2\over 12 M}\left[
(94+24(\mu_s+1)\tau_3)\log\left({m_\pi \over \mu}\right)+79+(16+12\mu_s)\tau_3\right]\right.\nonumber\\
&& \qquad \qquad \quad\left. -\frac 2 3 
 \left(c_2 - \frac{g_A^2}{8M}\right) \left(2\log\left({m_\pi \over \mu}\right)+1\right)\right\};\\
&& 4\pi\beta^{(4)}={e^2 \over 16\pi^2f_\pi^2}\left\{{g_A^2 \over 24 M}\left[
(94+48\mu_s\tau_3)\log\left({m_\pi \over \mu}\right)+51+24\mu_s\tau_3 \right]
\right.\nonumber\\
&& \qquad \qquad \left.   -\frac 2 3
\left(c_2 - \frac{g_A^2}{8M}\right) \log\left({m_\pi \over \mu}\right)-\frac 1 3 \left(c_2 - \frac{g_A^2}{8M} +2c_3-4c_1\right)\right\}\nonumber\\
&& \qquad \qquad \qquad + \delta \beta^s(\mu) + \tau_3 \delta \beta^v(\mu).
\label{eq:fourthorderpols}
\end{eqnarray}

The dependence on renormalization scale, $\mu$, in the isoscalar and isovector LECs $\delta \alpha^{s,v}$ and $\delta
\beta^{s,v}$ cancels that of the loops. These LECs encode
contributions to the polarizabilities from mechanisms other than soft-pion loops,
i.e. short-distance effects. In contrast, the spin polarizabilities, as
well as all higher terms in the Taylor-series expansion of the Compton
amplitude, are still predictive in $\chi$PT at $O(P^4)$. (The full amplitudes up to $O(P^4)$ are not written here; they
can be constructed from the details given in the Appendices of Ref.~\cite{Be04}.)

\begin{figure}[htb]
\begin{center}
\includegraphics[width=12cm]{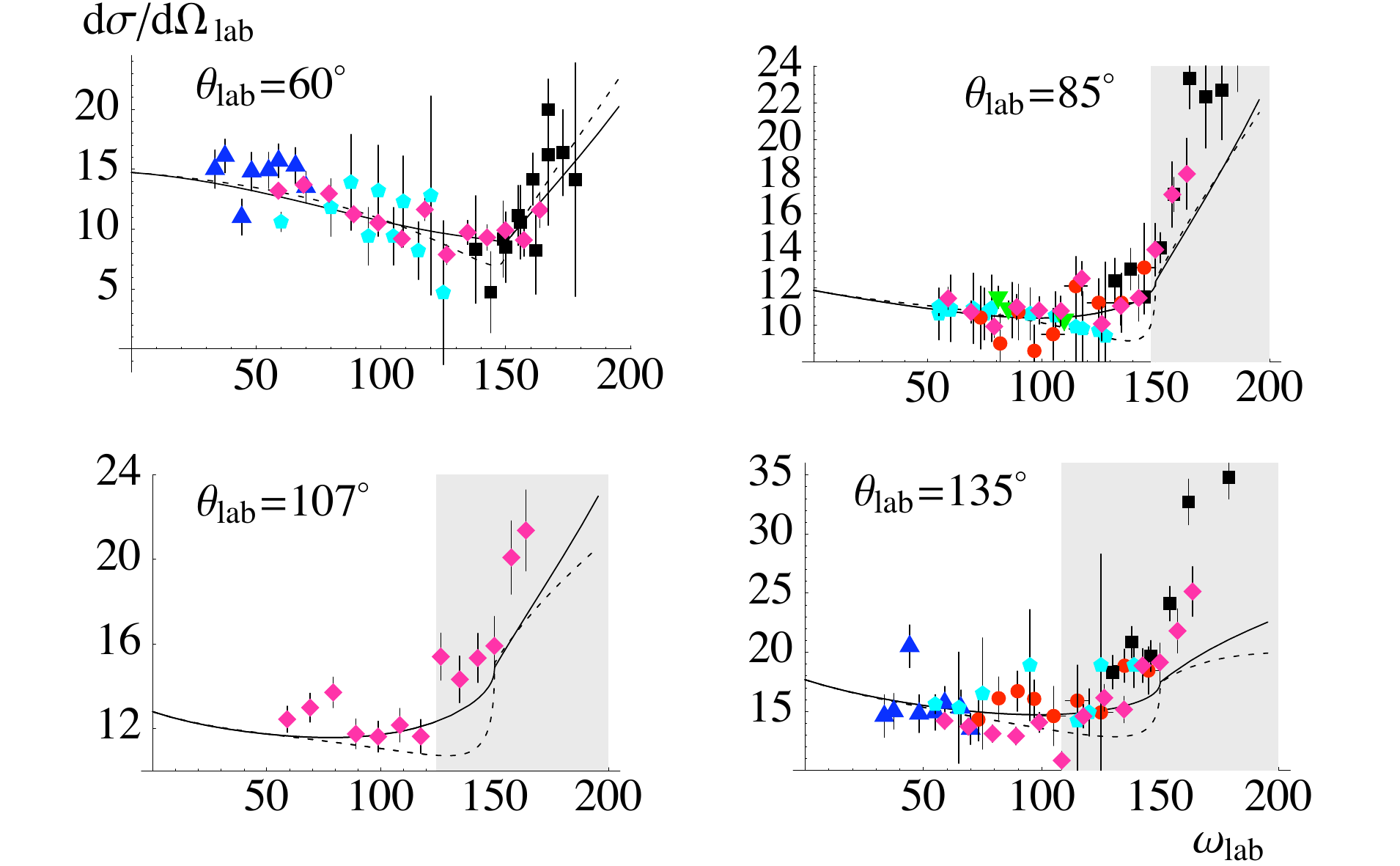}
\end{center}
\caption{\label{fig-protonplot} Results of the $O(Q^4)$ HB$\chi$PT best fit to
   the differential cross sections for Compton scattering on the
   proton at various angles, compared to the experimental data
   \protect\cite{pdata,mainz}. The gray region is excluded from the
   fit. The magenta diamonds are Mainz data \cite{mainz}; the other
   symbols are explained in Ref. \cite{McG01}. Figure reprinted from Ref.~\cite{Be04}, with permission from Elsevier.}
\end{figure}

Use of the $O(P^4)$ amplitude allows a good description of $\gamma$p data in the kinematic range $\omega,\sqrt{|t|} < 180$ MeV (see Fig.~\ref{fig-protonplot}). This calculation includes the effects (\ref{eq:As}), and also the additional $O(P^4)$ effects in Fig.~\ref{fig-protborn} and \ref{fig-protloop} that were computed in Ref.~\cite{McG01}. The only free parameters are $\alpha^p$ and $\beta^p$. A fit to data in this kinematic range yields~\cite{Be04,Be02}: 
\begin{eqnarray}
\alpha^p &=& (12.1 \pm 1.1)_{-0.5}^{+0.5} \times 10^{-4} \, {\rm fm}^3, 
\nonumber\\
\beta^p &=& (3.4 \pm 1.1)_{-0.1}^{+0.1} \times 10^{-4} \, {\rm fm}^3. 
\label{eq:protpol1}
\end{eqnarray}
Statistical (1-$\sigma$) errors for a two-parameter fit are inside the brackets, while the second set of errors are an estimate of the uncertainty due to the effect of higher-order terms. 

This success in $\gamma$p scattering prompts an important question about neutron polarizabilities: how different are they from those of the proton? In particular, the leading pion-cloud mechanisms in the first line of Fig.~\ref{fig-protloop} all yield exactly the same contributions for the neutron and the proton. So too do the dominant mechanisms associated with excitation of the Delta(1232)~\cite{He97,Bu91}. So, a measurement of a non-zero value of the isovector quantities $\alpha^v$ and $\beta^v$ (defined as in Eq.~(\ref{eq:fourthorderpols})) would be indicative of other (i.e. non-leading-Delta-excitation) short-distance effects in the polarizabilities.

In order to obtain information on neutron polarizabilities we must analyze Compton scattering from light nuclei. At photon energies $\sim m_\pi$ the approach to reactions described in Sec.~\ref{sec-chipt2N} tells us that the Compton amplitude for these nuclei is formed by computing 
\begin{equation}
T_{\gamma A}=\langle \psi_A|T_{\gamma N} + T_{\gamma NN}|\psi_A \rangle
\label{eq:wein}
\end{equation}
where $|\psi_A \rangle$ is the wave function of the nucleus, and $T_{\gamma N}$ is the single-nucleon Compton amplitude. $T_{\gamma NN}$ is the amplitude for $\gamma NN \rightarrow \gamma NN$, with all NN interactions after (before) the departure (arrival) of the outgoing (incoming) photon amputated from it. If these amplitudes are computed up to order $n$ in $\chi$PT, and the wave function $|\psi_A \rangle$ is obtained from a $\chi$PT potential that is also computed at that order, then all the consequences of chiral-symmetry breaking for the $\gamma$A amplitude have been accounted for---up to the $O(P^{n+1})$ corrections occurring at higher orders in the $\chi$ET expansion.

\begin{figure}[htbp]
\begin{center}
\includegraphics[width=12cm]{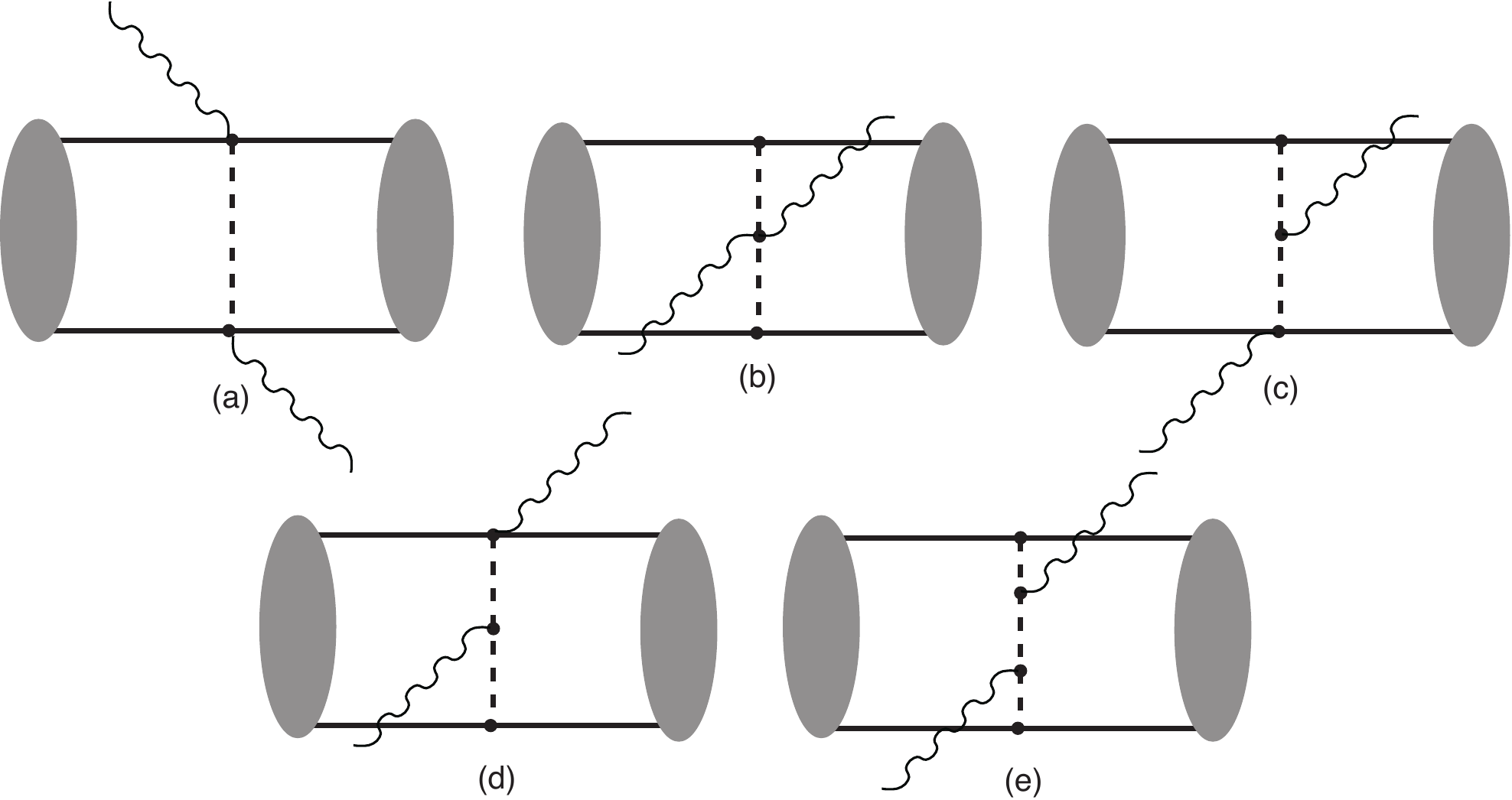} 
\end{center}
\caption {Two-body diagrams for the irreducible $\gamma NN \rightarrow \gamma NN$ kernel, $T_{\gamma NN}$, at 
$O(P^3)$, sandwiched between deuteron wave functions which are represented by the oval blobs. Permutations are not shown. Figure reprinted from Ref.~\cite{Be04}, with permission from Elsevier.
\vspace{0.3in} \label{fig-TgammaNN}}
\end{figure}

The $\chi$PT amplitude $T_{\gamma NN}$ was first computed in Ref.~\cite{Be99}, and includes the diagrams shown in Fig.~\ref{fig-TgammaNN}. At $O(P^3)$ it contains the two-nucleon analog of the mechanisms depicted in graphs (i)--(iv) of Fig.~\ref{fig-protloop}. These isoscalar exchange currents are sizeable at the $\sim 100$ MeV photon energies of interest to us here.

The $\chi$ET computation of $\gamma$d scattering has received extensive attention, with an $O(P^4)$ computation in Ref.~\cite{Be04,Be02}. That calculation showed that short-distance effects in the $\gamma$d reaction are not described with enough precision at $O(P^4)$ in $\chi$ET to allow an accurate extraction of polarizabilities from existing experimental data. The short-distance physics that is missing is of two types.

First, the Delta(1232) starts to have an impact beyond its effects on the ``static" polarizabilities, (\ref{eq:fourthorderpols}), already at $\omega \approx 80$ MeV---well below the upper energy of the $\gamma$d data set. This effect is particularly noticeable at backward angles, where magnetic scattering is known to lead to significant Delta excitation in the $\gamma$p case~\cite{PP03,Hi04,LP09}. Since the value of $\sqrt{|t|} \approx 150$ MeV for some existing $\gamma$d data, it is no surprise to find Delta(1232) effects that are nominally $O(P^6)$ in HB$\chi$PT have a $\approx 10$\% effect on the $\gamma$d differential cross section. This mars our attempts to describe $\gamma$d data in HB$\chi$PT at energies of order 100 MeV, and translates into $\sim 100$\% uncertainties in $\beta^s$. 

\begin{figure}[htbp]
\begin{center}
\includegraphics[width=6cm]{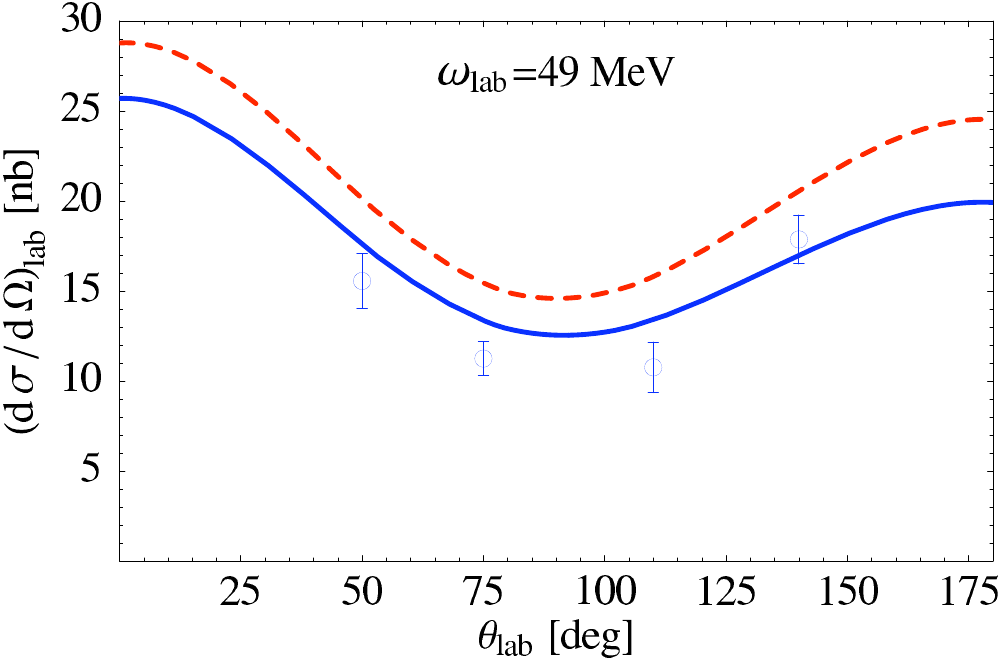} 
\includegraphics[width=6.2cm]{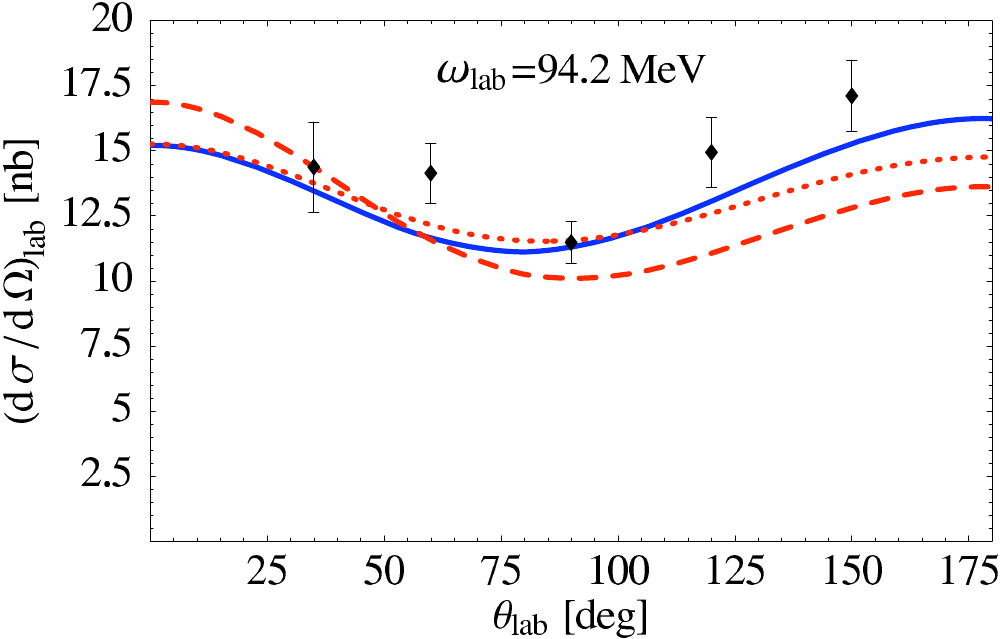}
\end{center}
\caption {Results for Compton scattering from the deuteron at photon lab. energies 49 MeV and 94.2 MeV~\cite{hg09}. Data shown are from Refs.~\cite{Lu94,Ho00} respectively. The long-dashed curve is an $O(P^3)$ calculation \'a la Ref.~\cite{Be99}, while the short-dashed curve is the result found in one of the fits of Ref.~\cite{Be04}. The solid line supplements the $O(P^3)$ result with the dominant effects of Delta(1232) excitation, and corresponds to the central values in Eq.~(\ref{eq:hildnos}). Figure courtesy H.~Grie\ss hammer.}
\label{fig-gammad}
\end{figure}

This deficiency was rectified in Ref.~\cite{Hi05a}, where the leading effects of Delta(1232) excitation were added to the $\gamma$d calculation. This requires the use of a new effective theory, in which the Delta(1232) is included as an explicit low-energy degree of freedom~\cite{He97,He98,Hi04}. That new EFT provides an accurate description of higher-energy data in both $\gamma$N and $\gamma$d scattering---as shown in Fig.~\ref{fig-gammad} for the $\gamma$d case. 

Second, there are short-distance, two-body Compton mechanisms that are suppressed by two powers of $P$, compared to the dominant $O(P^3)$ piece of $T_{\gamma NN}$. But, when $\chi$PT is applied to photon-nucleus scattering, the expansion parameter is the larger of $\sqrt{\omega/M}$ and $m_\pi/\sqrt{M \omega}$. Therefore these higher-order pieces can easily give a 10\% effect in the amplitude at 100 MeV. This was the size of differences seen in Ref.~\cite{Be04} when deuteron wave functions with different implementations of short-distance NN dynamics were used in the calculation. 

This difficulty can be circumvented by recognizing that in the regime of very low photon energies a different power counting is required. The power counting for $\omega \sim m_\pi^2/M$ was implemented in Ref.~\cite{Hi05b}. In that domain Eq.~(\ref{eq:wein}) is modified to:
\begin{equation}
T_{\gamma A}=\langle \psi_A|T^{irr}_{\gamma N} + T^{irr}_{\gamma NN} + J^\dagger G J + J G J^\dagger|\psi_A \rangle,
\label{eq:hild}
\end{equation}
where $J$ is the operator for photo-absorption by the two-nucleon system and $G$ is the fully interacting two-nucleon Green's function. In Eq.~(\ref{eq:hild}) the operator $T^{irr}_{\gamma N}$ includes only the one-nucleon-irreducible part of the $\gamma$N amplitudes, and a similar definition applies to $T^{irr}_{\gamma NN}$. This avoids double counting with the last two terms on the right-hand side of Eq.~(\ref{eq:hild}). If $J$, $T^{irr}_{\gamma N}$, $T^{irr}_{\gamma NN}$, and the NN potential $V$ are all computed up to the same order in $\chi$PT then the calculation of Eq.~(\ref{eq:hild}) is manifestly gauge invariant. Consequently, the correct Thomson limit is automatically obtained, and a significant reduction in the sensitivty of $T_{\gamma A}$ to details of the short-distance physics in the NN system results. 

This repairs the missing physics in the calculation of Ref.~\cite{Be04} because the part of $J^\dagger G J + J G J^\dagger$ that was {\it not} included in $T_{\gamma N}$ and $T_{\gamma NN}$ of Eq.~(\ref{eq:wein}), while higher-order in $\chi$ET, is not gauge invariant by itself. Once a gauge-invariant calculation is performed the short-distance NN dynamics can only impact the $\gamma A$ amplitude at $O(P^6)$ and beyond. The calculations of Ref.~\cite{Hi05b} show that these effects are very small,
with the uncertainty in the cross section due to the use of deuteron wave functions $|\psi_d \rangle$ that contain different short-distance physics being $< 2 \%$ in the experimentally relevant energy range. 

That range presently extends from $\omega \approx 45$ MeV to $\omega \approx 95$ MeV, and over angles from about $30^o$ to $150^o$. An extraction of neutron polarizabilities that is as accurate as that of Eq.~(\ref{eq:protpol1}) for the proton case therefore
requires the formalism of Ref.~\cite{Hi05b}, as well as some discussion of Delta(1232) effects. But, it is important to note that for $\omega=50$--$100$ MeV  the dominant effects in this computation arise from excitation of the pion cloud: the pion cloud of the nucleon that gives the dominant contribution to the electric polarizability, and the pion cloud of the deuteron that produces the exchange currents shown in Fig.~\ref{fig-TgammaNN}. Ref.~\cite{Hi05b} analyzed the data on elastic $\gamma$d scattering and extracted the result~\cite{hg09}:
\begin{eqnarray}
\alpha^s=(11.3 \pm 0.7 \pm 0.6 \pm 1.0) \times 10^{-4}~{\rm fm}^3, \nonumber\\
\beta^s=(2.8 \pm 0.7 \mp 0.6 \pm 1.0) \times 10^{-4}~{\rm fm}^3,
\label{eq:hildnos}
\end{eqnarray}
for the isoscalar polarizabilities. (The first error is the 1-$\sigma$ statistical error, the second comes from the use of the Baldin sum rule to constrain the fit, and the third is an estimate of the potential impact of omitted mechanisms on the result.) The quality of the fit is also good, see Fig.~\ref{fig-gammad} for examples at photon laboratory energies of 49 MeV and 94.2 MeV. The ongoing experiment at MAX-Lab in Lund, Sweden~\cite{LundCompton}, as well as future experiments at the High-Intensity Gamma-ray Source (HI$\gamma$S) at the Triangle Universities Nuclear Laboratory, will provide much-needed additional data to test this theoretical description, and improve the extraction of $\alpha^s$ and $\beta^s$.

Recently the first computation of elastic Compton scattering on the three-body system was completed~\cite{Ch06,Sh09}. This was also the first consistent $\chi$ET calculation of an electromagnetic process in the three-nucleon system. In this case the focus was on photon energies in the 60--120 MeV range, and the formula (\ref{eq:wein}) was used. This, then, was a first, exploratory, study that sought to examine what information Compton experiments using a Helium-3 target could give us about neutron polarizabilities. Once again the $T_{\gamma N}$ and $T_{\gamma NN}$ were computed up to $O(P^3)$, and a variety of potentials were employed in the calculation of $|\psi_{He^3} \rangle$.

The first conclusion from this study is that Helium-3 will have a significantly larger cross section for coherent Compton scattering than deuterium, due to the presence of two protons in the nucleus. Furthermore, the 
 Thomson terms of these two protons can interefere with the neutron polarizability pieces of the $\gamma$N amplitude, so these pieces have a larger effect than in $\gamma$d scattering---at least in absolute terms. 
 
 In addition, polarized Helium-3 is unique and interesting because
it seems to behave as an ``effective polarized neutron" in Compton scattering. The Compton response is---at least up to $O(P^3)$---dominated by the nuclear configuration where the two protons are paired in a relative S-wave, and the spin of the nucleus is carried by the neutron. And two-body currents at this order are largely spin independent. These two facts lead to $\gamma {}^3$He double-polarization observables that look very similar to the asymmetries that would be measured were the same double-polarization experiments done on a neutron target.

\begin{figure}[!htb]
\begin{center}
\includegraphics*[width=.37\linewidth]{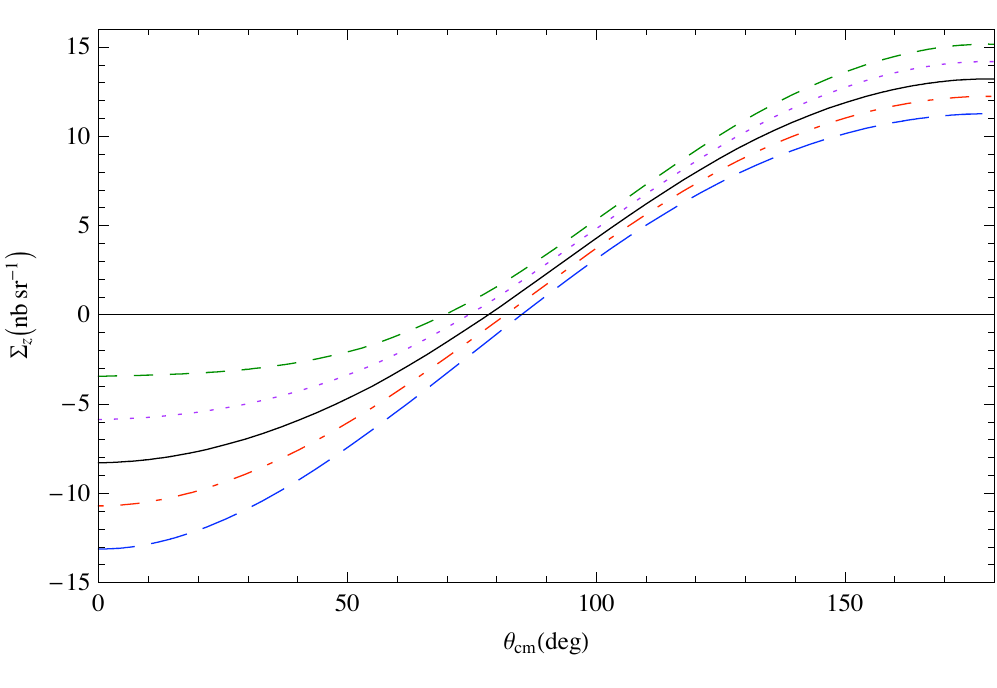}
\includegraphics*[width=.37\linewidth]{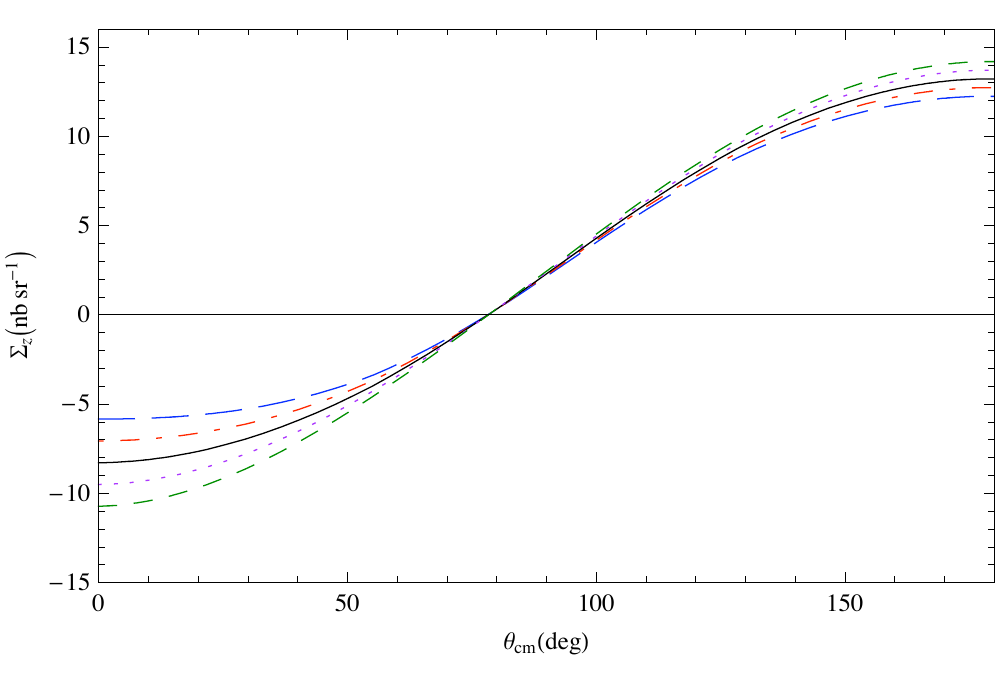}\\
\includegraphics*[width=.37\linewidth]{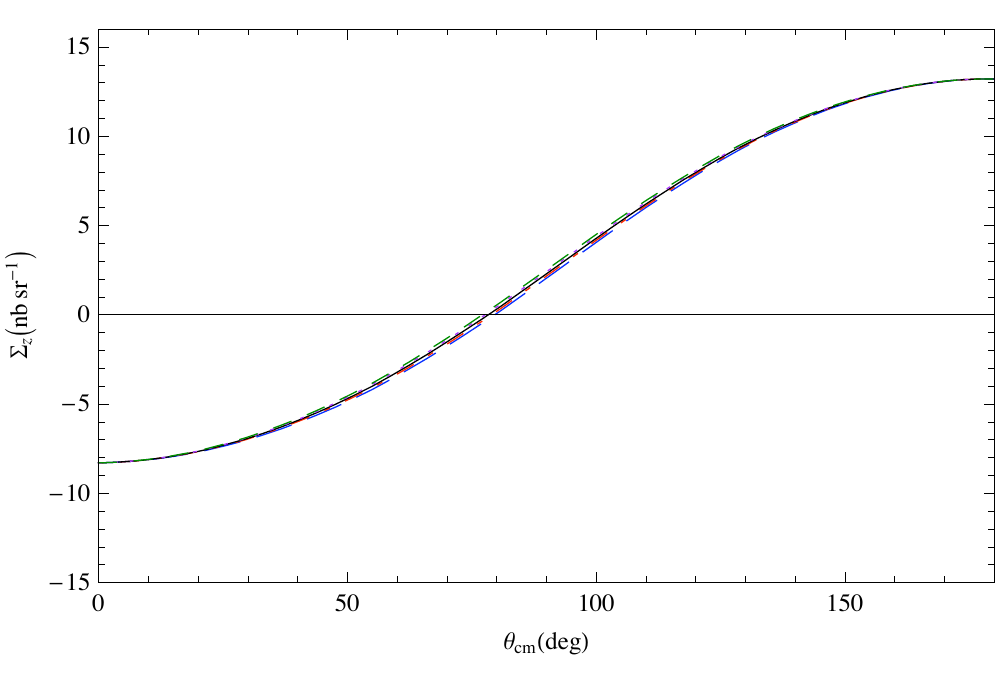}
\includegraphics*[width=.37\linewidth]{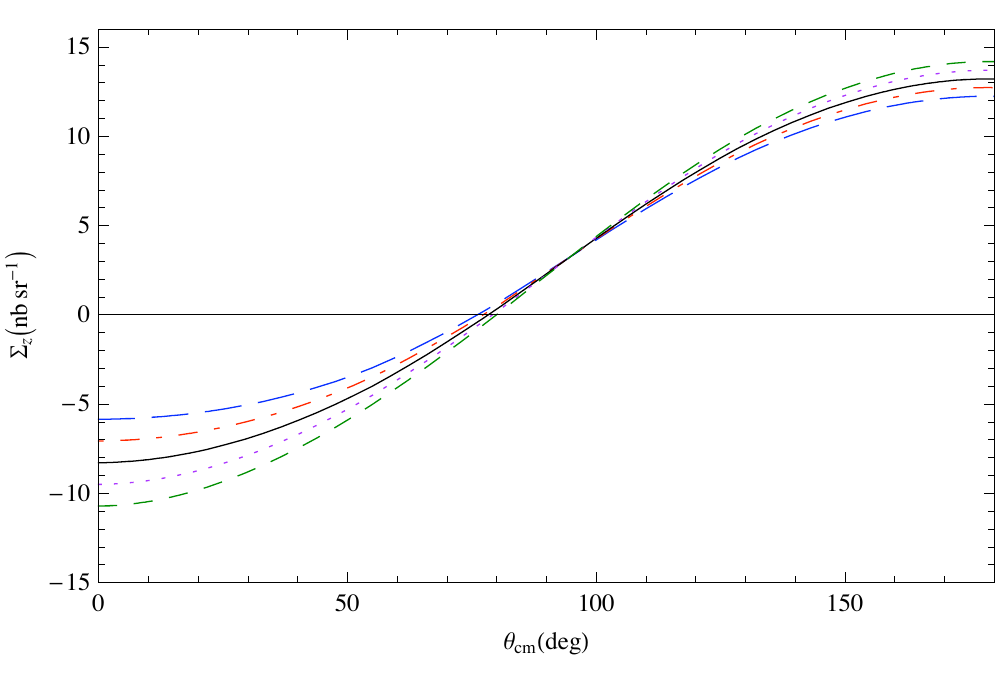}
\parbox{1.\textwidth}{
  \caption {The four panels above
correspond to the $O(P^3)$ result for $\Delta_z$, with an additional variation of, respectively, $\Delta \gamma_{1n}$
(top-left panel), $\Delta \gamma_{2n}$ (top-right panel), $\Delta \gamma_{3n}$
(bottom-left panel) and $\Delta \gamma_{4n}$ (bottom-right panel). The calculations are done in the c.m. frame at 120 MeV. The solid
(black) curves correspond to the full unperturbed $O(P^3)$ results. The
long-dashed (blue) curves correspond to $\Delta \gamma^n_{i} = - \gamma_i^n (O(P^3))$, dot-dashed (red) to $\Delta \gamma_i^n = - \gamma_{i}^n
 (O(P^3))/2$, dotted (magenta) to $\Delta \gamma_i^n = \gamma_i^n
 (O(P^3))/2$ and dashed (green) to $\Delta \gamma_i^n =
\gamma_{i n} (O(P^3)$. Figure reprinted from Ref.~\cite{Sh09}, with permission from Elsevier.}
  \label{fig:ddcsz}}
\end{center}
\end{figure}

Figure~\ref{fig:ddcsz} shows predictions for $\Delta_z$, which is defined as a difference of differential cross sections with circularly polarized photons and the target polarized parallel and anti-parallel to the beam direction, as a function of the scattering angle of 120 MeV photons. Here different neutron spin polarizabilities are implemented in an otherwise strict, $O(P^3)$, $\gamma$He$^3$ scattering calculation.
In each panel one of the four neutron spin polarizabilities is varied. This shows that a measurement of $\Delta_z$ at this energy should allow extraction of the combination $\gamma_1^n - (\gamma_2^n + 2 \gamma_4^n) \cos \theta$ (see Ref.~\cite{Sh09} for details).

\begin{figure}[!htb]
\begin{center}
\includegraphics*[width=.37\linewidth]{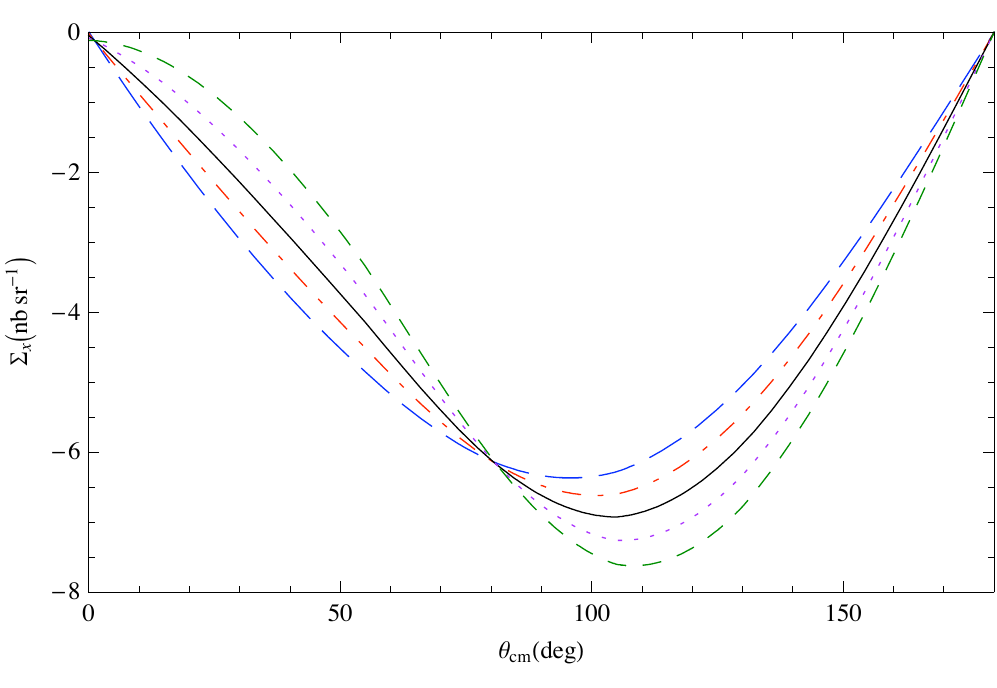}
\includegraphics*[width=.37\linewidth]{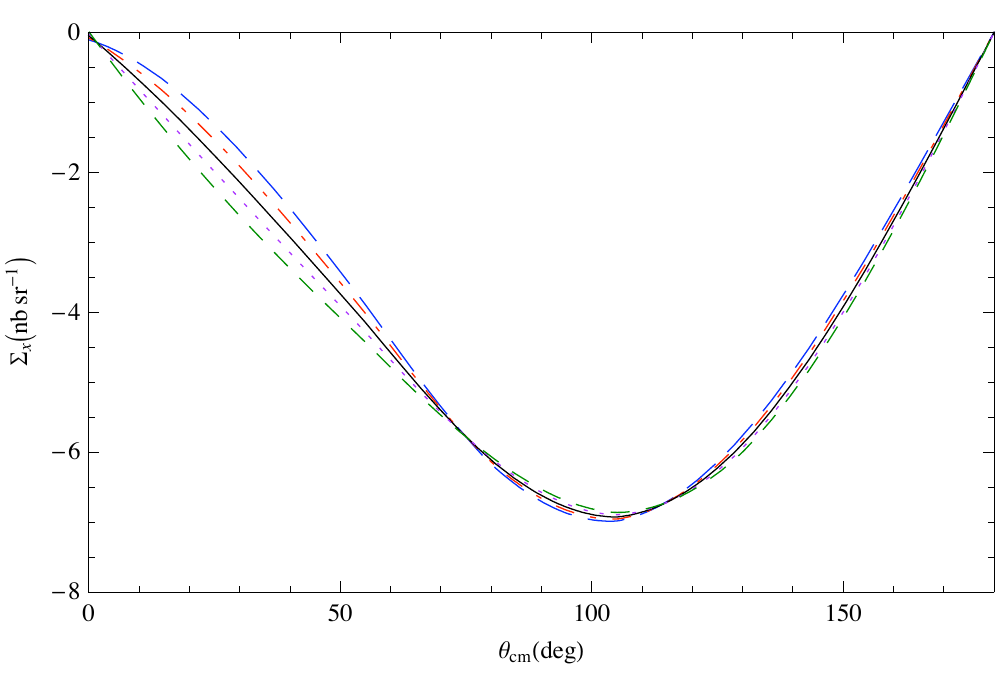}\\
\includegraphics*[width=.37\linewidth]{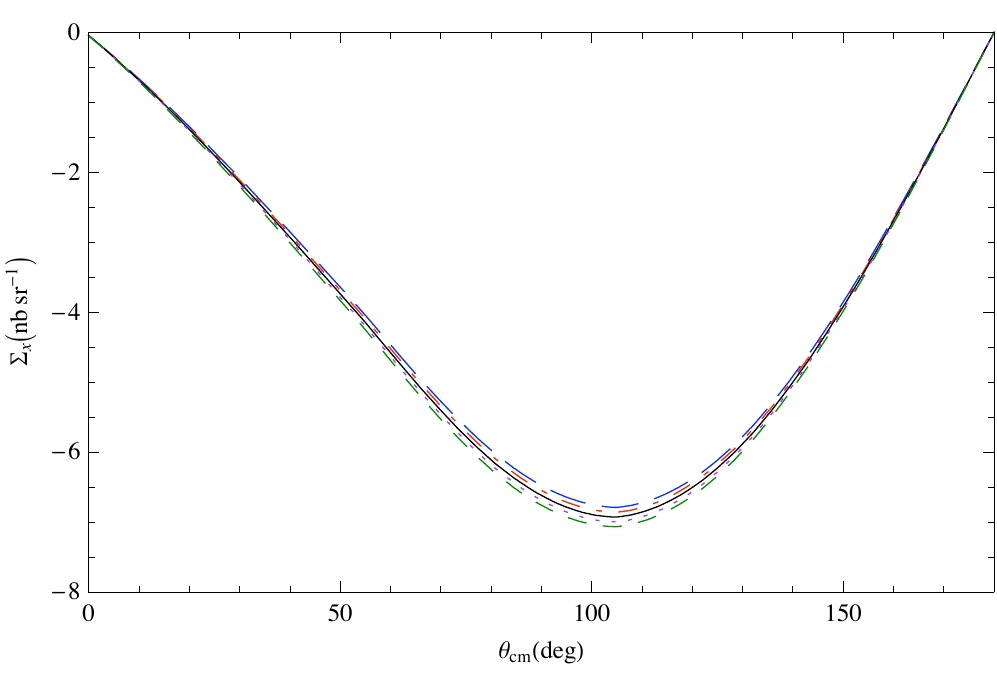}
\includegraphics*[width=.37\linewidth]{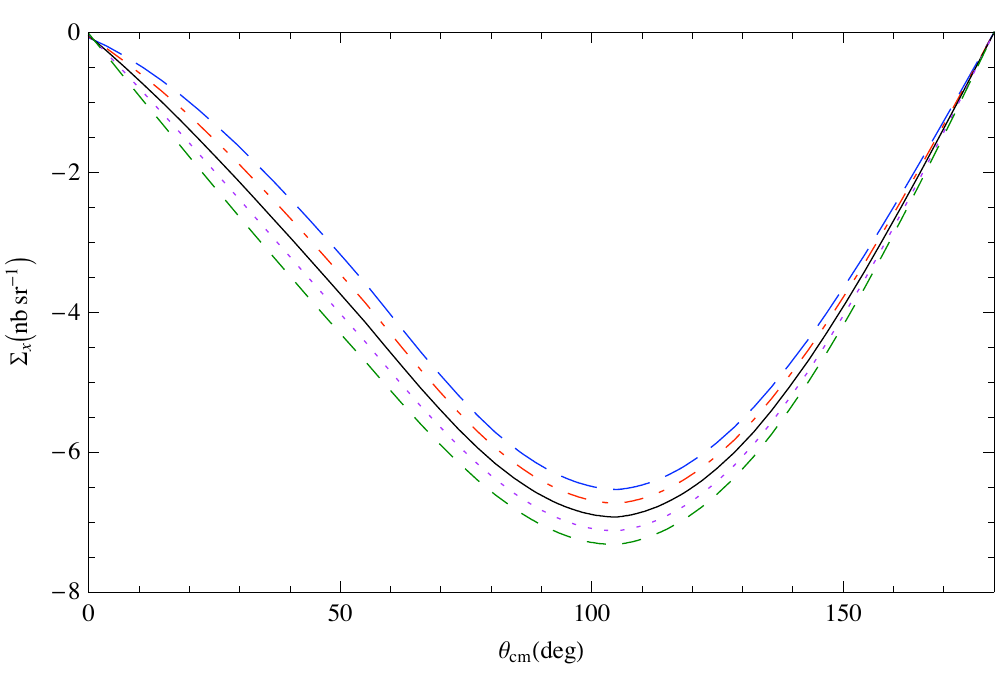}
\parbox{1.\textwidth}{
  \caption {The four panels above
display the $O(P^3)$ results for $\Delta_x$ at a c.m. photon energy of 120 MeV. Also shown are the results of varying one neutron spin polarizability per panel. Legend as in Fig.~\ref{fig:ddcsz}. Figure reprinted from Ref.~\cite{Sh09}, with permission from Elsevier.}
  \label{fig:ddcsx}}
\end{center}
\end{figure}

Next, in Fig.~\ref{fig:ddcsx} we plot $\Delta_x$ vs. the c.m. angle
at 120 MeV. Again, the different panels each correspond to varying one of the four
spin polarizabilities in an otherwise strict $O(P^3)$ $\chi$ET calculation. The results suggest that $\Delta_x$ is sensitive to a different combination of neutron spin polarizabilities than is $\Delta_z$.

Measurement of these obseravbles is a high priority for HI$\gamma$S, along with  polarized $\vec{\gamma} \vec {\rm p}$ and 
$\vec{\gamma} \vec {\rm d}$. The idea
is that the $\vec \gamma \vec p \rightarrow \gamma p$ experiments will
reveal proton spin polarizabilities, which can then be compared to the neutron ones measured in the $\vec{\gamma} \vec{{\rm He}^3}$ experiments. A different combination of neutron spin polarizabilities can be accessed through double-polarization measurements on the deuterium nucleus~\cite{CP05}.
There is also the possibility of coherent Helium-3 Compton
scattering measurements at MAX-Lab in the near future. This allows us to anticipate the measurement of 
several neutron spin polarizabilities, with the consequent possibility of examining differences between proton and neutron $\gamma_i$'s. There is also progress on the extraction of information on these quantities from lattice QCD~\cite{De08}, and so such experiments could provide important insights into the interplay of Goldstone-boson and higher-energy QCD dynamics in nucleon Compton scattering.

\section{Conclusion, and some words on other electromagnetic-induced reactions}

\label{sec-conc}

Electron scattering from the neutron provides the opportunity to map out its electromagnetic structure as a function of $Q^2$, the square of the space-like momentum transfer in the reaction. Neutron Compon scattering provides a different picture, as we probe neutron structure as a function of photon energy $\omega$. In both cases the rapid variation of response functions with the kinematic parameter is driven by the neutron's pion cloud. Slower variation comes from excitations of higher energy. The review of Ref.~\cite{DW08} places greater emphasis on the role of  those higher-energy degrees of freedom, and also covers a wider swath of kinematical territory, than does this article. The understanding of the nucleon's electromagnetic response I have presented here is based on the chiral symmetry of QCD, and the status of pions as Goldstone bosons of that symmetry. However, since chiral symmetry is only an approximate symmetry of the QCD Lagrangian, the ``slow" variation from more massive degrees of freedom can, on occasion, be just as fast as that due to pions, see e.g. the problem of the magnetic response of the nucleon in Compton scattering~\cite{Hi04,Bu91}. 

It is critical that we test and elucidate the consequences of QCD associated with this physics for both neutrons and protons. Chiral perturbation theory systematically implements the interplay of pion-cloud and shorter-distance mechanisms in nucleon structure. It predicts definite patterns for proton and neutron observables, e.g. that it is the isovector combination of form factors that has the dominant effects due to chiral symmetry, or that proton and neutron polarizabilities are equal at leading order. 
The calculations and experiments reported on here show that a picture of neutron structure based on chiral symmetry provides a good qualitative understanding of the neutron's electromagnetic excitation already at the leading one-loop level. That picture can be rendered quantitatively accurate if some care is taken to include the dominant short-distance mechanisms, e.g. those due to the Delta(1232), through use of the chiral effective field theories developed in 
Refs.~\cite{JM91,He97,He98,PP03}. 

An obvious extension of the tests discussed here is to examine the neutron's response as a function of $\omega$ and $Q^2$ {\it simultaneously}. This has been done for the proton, through $(e,e' \gamma)$ or ``virtual Compton scattering" (VCS) experiments. Chiral perturbation theory up to $O(P^3)$ provides a good description of the lowest-$Q^2$ experiment of this type~\cite{OOPS}. At this order, $\chi$PT not only correctly predicts the electric and magnetic polarizabilities of the proton to within experimental uncertainties, it also predicts the spatial distribution of the polarizability response within the proton~\cite{HemmertVCS}. The theory also predicts that the neutron's generalized polarizabilities are the same as the proton's at this order. Experiments to test this prediction would be very timely, and could be accomplished through a VCS experiment on the deuteron, together with a calculation of the two-body corrections to $O(P^3)$. More ambitiously, generalized neutron spin polarizabilities could perhaps be extracted from VCS experiments of the type recently undertaken at Mainz in the proton case~\cite{MainzVCS}, but with the proton replaced by a Helium-3 target. 

Of course, the most direct evidence for the chiral structure of the neutron comes from pion-production reactions. The amplitude for $\gamma n \rightarrow \pi^- p$ is given to very good accuracy by the leading chiral effect: the ``Kroll-Ruderman" term. In this case $O(P^3)$, where the leading-loop contributions appear, is two orders beyond leading and interesting short-distance effects also enter at this order~\cite{FHLU}. The $O(P^3)$ amplitude for charged-pion photoproduction on the neutron can be compared to experiment by examining data on pion capture by the proton---data which is in good agreement with the $\chi$PT result.  
An approved experiment at the MAX-Lab facility in Lund, Sweden, will soon measure the energy dependence of the near-threshold cross section for $\gamma d \rightarrow \pi^- p p$ to good accuracy~\cite{Fissum}. This should provide useful information on the charged-pion photoproduction reaction on the neutron, although again, this one-body part of the amplitude will have to be disentangled from NN mechanisms. 

Finally, one of the most beautiful pieces of evidence in neutron dynamics regarding the pattern of chiral-symmetry breaking occurs in the reaction $\gamma n \rightarrow \pi^0 n$. This is predicted to have a large threshold $E_{0+}$ amplitude, in spite of the absence of any neutron charge, because of the impact of the two-step process $\gamma n \rightarrow \pi^- p \rightarrow \pi^0 n$---even though the threshold for $\pi^0$ production is
below the charged-pion threshold. HB$\chi$PT predicts the neutron $E_{0+}$, based on the one-loop dynamics corresponding to this process, as well as the measured proton value, $E_{0+}^{\pi^0 p}$, and  the assumption that the short-distance physics in the proton and neutron reactions is the same  (apart from obvious isospin factors). The result---which includes terms up to $O(P^4)$ in HB$\chi$PT---is~\cite{Be96}:
\begin{equation}
E_{0+}^{\pi^0 n}=2.13 \times 10^{-3} m_\pi^{-1}.
\label{eq:E0pluspi0n}
\end{equation}

HB$\chi$PT's prediction for $\gamma n \rightarrow \pi^0 n$ can be tested in $\gamma d \rightarrow \pi^0 d$. But, a consistent calculation of the threshold S-wave amplitude for neutral-pion photoproduction on deuterium, $E_d$, requires the inclusion of the $O(P^3)$ and $O(P^4)$ mechanisms for $\gamma NN \rightarrow \pi^0 NN$~\cite{Be95,Be97}. These diagrams include the double-scattering contributions that are the two-body counterpart of the diagrams that in the single-nucleon case lead to the large neutron $E_{0+}$. $\chi$ET up to $O(P^4)$ then predicts:\begin{equation}
E_d=(-2.6 \pm 0.2) \times 10^{-3} m_\pi^{-1} + 0.38 E_{0+}^{\pi^0 n}.
\label{eq:gammadpi0dth}
\end{equation}
The uncertainty comes from the experimental error on $E_{0+}^{\pi^0 p}$ and the variation in the result with different NN-system wave functions. (Note there are other higher-order uncertainties that are {\it not} included in this error bar.) Using a sequence of measurements at energies up to 20 MeV above threshold~\cite{Bergstrom} to obtain $E_d$, and then using the result of Eq.~(\ref{eq:gammadpi0dth}) to convert this to a result for the threshold $\pi^0 n$ amplitude, we find:
\begin{equation}
E_{0+}^n=(3.03 \pm 0.58) \times 10^{-3} m_\pi^{-1},
\label{eq:gammadpi0dexpt}
\end{equation}
with an additional, here unquantified, uncertainty arising from the subtraction of the inelastic channel that was employed in Ref.~\cite{Bergstrom}.
The reasonable level of agreement between the $O(P^4)$ prediction (\ref{eq:E0pluspi0n}) and the measurement (\ref{eq:gammadpi0dexpt}) shows that we understand the chiral dynamics of the neutron, as revealed in $\gamma d \rightarrow \pi^0 d$ at threshold, quite well. 

Unfortunately the situation in $e d \rightarrow e' \pi^0 d$ is not quite as clear~\cite{Krebs}. For the case of threshold pions and low momentum transfer, a successful description could reasonably have been expected, since  the pion cloud should also dominate this process. However, there are puzzling discrepanices between theory and data already for $Q^2=0.1$ GeV$^2$. It is hoped that careful treatment of the NN system dynamics, together with a thorough examination of the Delta(1232) mechanisms that enter this process, can explain why the pion cloud dominates $\gamma d \rightarrow \pi^0 d$, but is in competition with higher-energy QCD dynamics in the case of electro-production. 

Studies that explore these issues will ensure that photo- and electro-production continue to provide important windows on the chiral structure of the neutron. A unified picture of low-energy and low-momentum-transfer neutron structure as revealed in these reactions, as well as neutron Compton scattering (real and virtual), and the neutron electromagnetic form factors has been established. This picture---which is systematized by chiral perturbation theory---is grounded in QCD and agrees with data in all these reactions at the expected level given the $\chi$PT order to which calculations have been accomplished thus far.
Ongoing efforts in both theory and experiment---and continued dialogue between them---will elucidate further details of neutron structure, and so improve our understanding of QCD in its strongly-interacting regime.

\section*{Acknowledgements}
This work was supported in part by the U.~S.  Department of Energy, Office of Nuclear Physics,
under contract No. DE-FG02-93ER40756 with Ohio University. I thank Matthias Schindler for a careful reading of---and useful comments on---a draft of this manuscript, as well as Hans-Werner Hammer for help in preparing Fig.~\ref{fig-neutronffs}, and J\"org Friedrich for making his database of neutron form-factor measurements available to me. I also thank 
numerous co-workers and interlocutors for enjoyable collaborations and discussions on the topics covered by this article. I am grateful to the Centre for the Subatomic Structure of Matter at the University of Adelaide and the Nuclear Theory Group at the University of Washington
for hospitality during the writing of this article.

\end{document}